  \documentclass[12pt]{article}
 \usepackage{ifpdf}
 \usepackage{epsfig}
  \usepackage[latin1]{inputenc}
 \usepackage{amsfonts}
 \usepackage{amsthm}
 \usepackage{amssymb, latexsym, amsmath}
 \usepackage{amsbsy}
\usepackage{amstext}
\usepackage{amsopn}
\usepackage{amsgen}
\usepackage[dvips]{color}
 \usepackage{graphicx}
  \usepackage{color}
 \newcommand{\inc}{{\it i}}

 \newcommand{\be}{\begin{equation}}
 \newcommand{\ee}{\end{equation}}
 \newcommand{\ba}{\begin{eqnarray}}
 \newcommand{\ea}{\end{eqnarray}}

 \newcommand{\erbold}{\mbox{{\boldmath $\vec r$}}}

  \newcommand{\eRbold}{\mbox{{\boldmath $\vec{R}$}}}
  \newcommand{\Rbold}{\mbox{{\boldmath $\vec{R}$}}}

\begin{document}
 \title{
                 ${{~~~~~~~~~~~~~}^{^{^{^{
   The~Astrophysical~Journal
   ,\,~746~:~150\,.~~~/\,2~February~2012\,/
                  }}}}}$
                  ~\\
 {\Large{\textbf{Tidal dissipation compared to seismic dissipation:\\ in small bodies, in earths, and in superearths
 \\}
            }}}
 \author{
  {\Large{Michael Efroimsky}}\\
  {\small{US Naval Observatory, Washington DC 20392 USA}}\\
  {\small{e-mail: ~michael.efroimsky @ usno.navy.mil~}}
  }
 \date{}

 \maketitle
 \begin{abstract}
 While the seismic quality factor and phase lag are defined solely by the bulk properties of the mantle, their tidal counterparts are
 determined both by the bulk properties and the size effect (self-gravitation of a body as a whole). For a qualitative estimate, we
 model the body with a homogeneous sphere, and express the tidal phase lag through the lag in a sample of material. Although simplistic,
 our model is sufficient to understand that the lags are not identical. The difference emerges because self-gravitation pulls the tidal
 bulge down. At low frequencies, this reduces strain and the damping rate, and makes tidal damping less efficient in larger objects.
 At higher frequencies, competition between self-gravitation and rheology becomes more complex, though for sufficiently large
 superearths the same rule applies: the larger the planet, the weaker tidal dissipation in it. Being negligible
 for small terrestrial planets and moons, the difference between the seismic and tidal lagging (and likewise between the seismic and
 tidal damping) becomes very considerable for large exoplanets (superearths).
 In those, it is much lower than what one might expect from using a seismic quality factor.

 The tidal damping rate deviates from the seismic damping rate especially in the zero-frequency limit, and this difference takes place
 for bodies of {\it{any}} size. So the equal in magnitude but opposite in sign tidal torques, exerted on one another by the primary and
 the secondary, have their orbital averages going smoothly through zero as the secondary crosses the synchronous orbit.

 We describe the mantle rheology with the Andrade model, allowing it to lean toward the Maxwell model at the lowest frequencies. To
 implement this additional flexibility, we reformulate the Andrade model by endowing it with a free parameter $\,\zeta\,$ which is the
 ratio of the Andrade timescale to the viscoelastic Maxwell time of the mantle. Some uncertainty in this parameter's
 frequency dependence does not influence our principal conclusions.

 \end{abstract}

 \section{The goal and the plan}

 As the research on exoplanetary systems is gaining momentum, more and more accurate theoretical tools of planetary dynamics come into
 demand. Among those tools are the methods of calculation of tidal evolution of both orbital and rotational motion of planets and their
 moons. Such calculations involve two kind of integral parameters of celestial bodies -- the Love numbers and the tidal quality factors.
 The values of these parameters depend upon the rheology of a body, as well as its size, temperature, and the tidal frequency.

 It has recently become almost conventional in the literature to assume that the tidal quality factor of superearths should be of
 the order of one hundred to several hundred (Carter et al. 2011, Léger et al. 2009). Although an acceptable estimate for the
 seismic $\,Q\,$, this range of numbers turns out to fall short, sometimes by orders of magnitude, of the tidal $\,Q\,$ of
 superearths.

 In our paper, the frequency dependence of tidal damping in a near-spherical homogeneous body is juxtaposed with the
 frequency dependence of damping in a sample of the material of which the body consists. For brevity, damping
 in a sample will be termed (somewhat broadly) as ``seismic damping".

 We shall demonstrate that, while the tidal $\,Q\,$ of the solid Earth happens not to deviate much from the solid-Earth seismic
 $\,Q\,$, the situation with larger telluric bodies is considerably different. The difference stems from the presence of
 self-gravitation, which suppresses the tidal bulge and thereby acts as extra rigidity -- a long-known circumstance often neglected
 in astronomical studies.\footnote{~Including the size effect via $\,k_l\,$ is common. Unfortunately, it is commonly assumed
 sufficient. This treatment however is inconsistent in that it ignores the inseparable connection between the Love number and the
 tidal quality factor (or the tidal phase lag). In reality, both the Love number and the sine of the tidal lag should be derived
 from the rheology and geometry of the celestial body, and cannot be adjusted separately from one another.} Due to self-gravitation
 (``size effect"), tidal damping in superearths is much less efficient than in earths, and the difference may come to orders of
 magnitude, as will be demonstrated below. Thus, while the {\it{seismic}} $\,Q\,$ of a superearth may be comparable to the seismic
 $\,Q\,$ of the solid Earth, the {\it{tidal}} $\,Q\,$ of a superearth may exceed this superearth's seismic $\,Q\,$ greatly. This is
 the reason why it is inappropriate to approximate superearths' tidal quality factors with that of the solid Earth.

 We shall show that the difference between the frequency dependence of the tidal $\,Q\,$ factor and that of the seismic $\,Q\,$ may
 explain the ``improper" frequency dependence of the tidal dissipation rate measured by the lunar laser ranging (LLR) method. We also shall point
 out that the correct frequency dependence of the tidal dissipation rate, especially at low frequencies, plays an important role in
 modeling the process of entrapment into spin-orbit resonances. In greater detail, the latter circumstance will be discussed in
 Efroimsky (2012).

 The rate of the ``seismic damping" (a term that we employ to denote also damping in a sample of the material) is defined, at each
 frequency, by the material's rheology only, i.e., by the constitutive equation linking the strain and stress at this frequency. The
 rate of the tidal damping however is determined both by the rheology and by the intensity of self-gravitation of the body. At a
 qualitative level, this can be illustrated by the presence of two terms, $\,1\,$ and $\,19\mu(\infty)/(2\rho\mbox{g}R)\,$, in the
 denominator of the expression for the static Love number $\,k_2\,$ of a homogeneous sphere. Here $\,\mu(\infty)\,$ denotes the relaxed
 shear modulus, g signifies the surface gravity, while $\rho$ and $\,R\,$ stand for the mean density and the radius of the body. The
 first of these terms, $\,1\,$, is responsible for the size effect (self-gravitation), the second for the bulk properties of the medium.
 Within the applicability realm of an important theorem called {\it{elastic-viscoelastic analogy}} (also referred to as the
 {\it{correspondence principle}}), the same expression interconnects the Fourier component $\,\bar{k}_2(\chi)\,$ of the time derivative
 of the Love number with the Fourier component $\,\bar{\mu}(\chi)\,$ of the stress-relaxation function at
 frequency $\,\chi\,$. This renders the frequency-dependence of the tangent of the tidal lag, which is the negative ratio of the
 imaginary and real parts of $\,\bar{k}_2(\chi)\,$.

 This preliminary consideration illustrates the way rheology enters the picture. First, the constitutive equation defines the
 frequency dependence of the complex compliance, $\,\bar{J}(\chi)\,$, and of the complex rigidity $\,\bar{\mu}(\chi)=
 1/\bar{J}(\chi)\,$. The functional form of this dependence determines the frequency dependence of the complex Love number, $\,\bar{k}_{
 \it{l}}(\chi)\,$. The latter furnishes the frequency dependence of the products $\,|\bar{k}_{l}(\chi)|\;\sin\epsilon_l(\chi)\,$ which
 enter the tidal theory.

 In Section \ref{formalism}, we briefly recall the standard description of stress-strain relaxation and dissipation in linear media.
 In Section \ref{Andrade_section}, we describe a rheological model, which has proven to be adequate to the experimental data on the
 mantle minerals and partial melts. The goal of the subsequent sections will be to build the rheology into the theory of bodily tides,
 and to compare a tidal response of a near-spherical body to a seismic response rendered by the medium. Finally, several examples will
 be provided. Among these, will be the case of the Moon, whose ``improper" tidal-dissipation frequency-dependence finds an explanation
 as soon as the difference between the seismic and tidal friction is brought to light. In the closing section, we shall compare our
 results with those obtained by Goldreich (1963).

 \section{Formalism}\label{formalism}

 Everywhere in this paper we shall take into consideration only the deviatoric stresses and strains, thus neglecting compressibility.

  \subsection{Compliance and rigidity.\\The standard linear formalism in the time domain}

 The value of strain in a material depends only on the present and past values taken by the stress and not on the current
 {\it{rate}} of change of the stress.
 Hence the compliance operator $\,\hat{J}\,$ mapping the stress $\,\sigma_{\gamma\nu}\,$ to the strain $\,u_{\gamma\nu}\,$ must be just
 an integral operator, linear at small deformations:
 \ba
 \label{L18}
 \label{subequations_1}
 2\,u_{\gamma\nu}(t)\;=\;\hat{J}(t)\;\sigma_{\gamma\nu}
 \;=\;\int^{t}_{-\infty}J(t\,-\,t\,')\;\stackrel{\centerdot}{\sigma}_{\gamma\nu}(t\,')\;dt\,'\;\;\;,
 \;\;\;
 \label{subequations_1_a}
 \label{18a}
 \ea
 where $\,t\,' < t\,$, while overdot denotes $\,d/dt\,'\,$.
 The kernel $\,J(t-t\,')\,$ is termed the {\it{compliance function}} or the {\it{creep-response function}}.

 Integration by parts renders:
 \ba
 2\,u_{\gamma\nu}(t)\;=\;\hat{J}(t)~\sigma_{\gamma\nu} ~=~
 J(0)\;\sigma_{\gamma\nu}(t)\;-\;J(\infty)\;\sigma_{\gamma\nu}(-\infty)
 ~+~\int^{t}_{-\infty}\stackrel{\;\centerdot}{J}(t\,-\,t\,')~{\sigma}_{\gamma\nu}(t\,')~d t\,'~~\,.~~~~\,
 \label{I12_3}
 \ea
 As the load in the infinite past may be set zero, the term containing the relaxed compliance $\,J(\infty)\,$ may be dropped. The unrelaxed
 compliance $\,J(0)\,$ can be absorbed into the integral if we agree that the elastic contribution enters the compliance function not as
 \ba
 J(t-t\,')\,=\,J(0)\,+\,\mbox{viscous and hereditary terms}~~~,
 \label{conv}
 \ea
 but as
 \ba
 J(t-t\,')\,=\,J(0)\,\Theta(t\,-\,t\,')\,+\,\mbox{viscous and hereditary terms}~~~.
 \label{convention}
 \ea
 The Heaviside step-function $\,\Theta(t\,-\,t\,')\,$ is set unity for $\,t-t\,'\,\geq\,0\,$, and zero for $\,t-t\,'\,<\,0\,$,
 so its derivative is the delta-function $\,\delta(t\,-\,t\,')\,$. Keeping this in mind, we reshape (\ref{I12_3}) into
 \ba
 2\,u_{\gamma\nu}(t)\,=\,\hat{J}(t)~\sigma_{\gamma\nu}\,=\,\int^{t}_{-\infty}\stackrel{\;\centerdot}{J}(t-t\,')~
 {\sigma}_{\gamma\nu}(t\,')\,d t\,'~~,~~~\mbox{with}~~
 J(t-t\,')~~\mbox{containing}~~J(0)\,\Theta(t-t\,')~~.~~~
 \label{I12_4}
 \ea

 Inverse to the compliance operator
 \ba
 2\,u_{\gamma\nu}\;=\;\hat{J}~\sigma_{\gamma\nu}~~~.
 \label{}
 \ea
 is the rigidity operator
 \ba
 \sigma_{\gamma\nu}\;=\;2\,\hat{\mu}~u_{\gamma\nu}~~~.
 \label{kera}
 \ea
 In the presence of viscosity, operator $\,\hat{\mu}\,$ is not integral but is integrodifferential, and thus cannot be expressed as
 $~\sigma_{\gamma\nu}(t)\,=\,2\,\int_{-\infty}^{t}\,\dot{\mu}(t\,-\,t\,')\,u_{\gamma\nu}(t\,')\,dt\,'~$. It can though be written as
 \ba
 \sigma_{\gamma\nu}(t)\,=\,2\,\int_{-\infty}^{t}\,{\mu}(t\,-\,t\,')\,\dot{u}_{\gamma\nu}(t\,')\,dt\,'\;\;\;,
 \label{permitted_1}
 \ea
 if its kernel, the stress-relaxation function $\,{\mu}(t\,-\,t\,')\,$, is imparted with a term $\,\,2\,\eta\,\delta(t-t\,')\,$,
 integration whereof renders the viscous portion of stress, $~2\,\eta\,\dot{u}_{\gamma\nu}\,$. The kernel also incorporates an
 unrelaxed part $\,\mu(0)\,\Theta(t\,-\,t\,')\,$, whose integration furnishes the elastic portion of the stress. The unrelaxed rigidity
 $\,\mu(0)\,$ is inverse to the unrelaxed compliance $\,J(0)\,$.

 Each term in $\,\mu(t-t\,')\,$, which is neither constant nor proportional to a delta function, is responsible for hereditary reaction.

  \subsection{In the frequency domain}

 To Fourier-expand a real function, nonnegative frequencies are sufficient. Thus we write:
  \ba
 \sigma_{\gamma\nu}(t)~=~\int_{0}^{\infty}\,\bar{\sigma}_{\gamma\nu}(\chi)~e^{\textstyle{^{\,\inc\chi t}}}~d\chi\quad\quad\mbox{and}
 ~\quad~\quad
 u_{\gamma\nu}(t)~=~\int_{0}^{\infty}\,\bar{u}_{\gamma\nu}(\chi)~e^{\textstyle{^{\,\inc\chi t}}}~d\chi~~~,
 \label{phys}
 \ea
 where the complex amplitudes are
 \ba
 {\bar{{\sigma}}_{\gamma\nu}}(\chi)={{{\sigma}}_{\gamma\nu}}(\chi)\,\;e^{\inc\varphi_\sigma(\chi)}~~~~~,
 ~~~~~~{\bar{{u}}_{\gamma\nu}}(\chi)={{{u}}_{\gamma\nu}}(\chi)\,\;e^{\inc\varphi_u(\chi)}~~~,
 \label{compamp}
 \label{L17}
 \ea
 while the initial phases $\,\varphi_{\sigma}(\chi)\,$ and $\,\varphi_{u}(\chi)\,$ are set to render the real amplitudes
 $\,\sigma_{\gamma\nu}(\chi_{\textstyle{_n}})\,$ and $\,u_{\gamma\nu}(\chi_{\textstyle{_n}})\,$ non-negative.
  To ensure convergence, the frequency is, whenever necessary, assumed to approach the real axis from below: $\,{\cal I}{\it{m}}
 (\chi)\rightarrow 0-\;\,$.

 With the same caveats, the complex compliance $\,\bar{J}(\chi)\,$ is introduced as the Fourier image of the time derivative of the
 creep-response function:
 \ba
 \label{L20}
 \int_{0}^{\infty}\bar{J}(\chi)\,e^{\inc\chi \tau}d\chi\,=\,\stackrel{\;\centerdot}{J}(\tau)~~.
 \ea
 The inverse expression,
 \ba
 \bar{J}(\chi)~=\,\int_{0}^{\infty}\stackrel{\;\centerdot}{J}(\tau)\,e^{-\inc\chi \tau}\,d\tau~~~,
 \label{inverse}
 \ea
 is often written down as
 \ba
 \bar{J}(\chi)~=\,J(0)\,+\,\inc\,\chi\,\int_{0}^{\infty}\left[\;J(\tau)\,-\,J(0)\,\Theta(\tau)\;\right]\;e^{-\inc\chi \tau}\,d\tau
              ~~\,.\,~~~~
 \ea
 For causality reasons, the integration over $\tau$ spans the interval $\,\left[\right.0,\infty\left.\right)\,$ only. Alternatively, we
 can accept the convention that {\it{each}} term in the creep-response function is accompanied with the Heaviside step function.

 Insertion of the Fourier integrals (\ref{phys} - \ref{L20}) into (\ref{L18}) leads us to
 \ba
 2\,\int_{0}^{\infty}\bar{u}_{\gamma\nu}(\chi)~e^{\textstyle{^{\,\inc\chi t}}}~d\chi
 \;=\;\int_{0}^{\infty}\bar{\sigma}_{\mu\nu}(\chi)~\bar{J}(\chi)~e^{\textstyle{^{\,\inc\chi t}}}~d\chi
  ~~~,
 \label{strain}
 \label{11strainn}
 \ea
 whence we obtain:
 \ba
 2\;\bar{u}_{\gamma\nu}(\chi)\,=\;\bar{J}(\chi)\;\bar{\sigma}_{\gamma\nu}(\chi)\;\;\;.
 \label{LLJJKK}
 \ea
 Expressing the complex compliance as
 \ba
 \bar{J}(\chi)\;=\;|\bar{J}(\chi)|\;\exp\left[\,-\,\inc\,\delta(\chi)\,\right]\;\;\;,\;
 \label{L25}
 \ea
 where
 \ba
 \tan \delta(\chi)\;\equiv\;-\;\frac{\cal{I}\it{m}\left[ \,\bar{J\,}(\chi)\, \right]}{\cal{R}\it{e}\left[ \,\bar{J\,}(\chi)\,\right]}
 \;\;\;,
 \label{delta_def}
 \label{L28}
 \ea
 we see that $\;\delta(\chi)\,\;$ is the phase lag of a strain harmonic mode relative to the
 appropriate harmonic mode of the stress:
 \ba
 \varphi_u(\chi)\;=\;\varphi_{\sigma}(\chi)\;-\;\delta(\chi)\;\;\;.
 \label{L27}
 \ea

 \subsection{The quality factor(s)}\label{damp}

 In the linear approximation, at each frequency $\,\chi\,$ the average (per period) energy dissipation rate $\langle\stackrel{
 \centerdot}{E}(\chi)\rangle$ is defined by the deformation at that frequency only, and bears no dependence upon the other frequencies:
 \ba
 \langle\,\dot{E}(\chi)\,\rangle \;=\;-\;\frac{\textstyle\chi E_{peak}(\chi)}{\textstyle Q(\chi)}\;
 \label{dissipation_rate}
 \label{L11}
 \ea
 or, the same:
 \ba
 \Delta E_{cycle}(\chi)\;=\;-\;\frac{2\;\pi\;E_{peak}(\chi)}{Q(\chi)}\;\;\;,
 \label{dega}
 \label{L12}
 \ea
 $\Delta E_{cycle}(\chi)\,$ being the one-cycle energy loss, and $\,Q(\chi)\,$ being the quality factor related to the phase lag at the
 frequency $\,\chi\,$. It should be clarified right away, to which of the lags we are linking the quality factor. When we are talking
 about a sample of material, this lag is simply $\,\delta(\chi)\,$ introduced above as the negative argument of the appropriate Fourier
 component of the complex compliance -- see formulae (\ref{L28} - \ref{L27}). However, whenever we address tide, the quality factor
 becomes linked (via the same formulae) to the {\it{tidal}} phase lag $\,\epsilon(\chi)\,$. Within the same rheological model, the
 expression for $\,\epsilon(\chi)\,$ differs from that for $\,\delta(\chi)\,$, because the tidal lag depends not only upon the local
 properties of the material, but also upon self-gravitation of the body as a whole.

 The aforementioned ``seismic-or-tidal" ambiguity in definition of $\,Q\,$ becomes curable as soon as one points out to which kind
 of deformation the quality factor pertains. More serious is the ambiguity stemming from the freedom in defining $\,E_{peak}(\chi)\,$.

 If $\,E_{peak}(\chi)\,$ in (\ref{L11} - \ref{L12}) signifies the peak $\,${\it{energy}}$\,$ stored at frequency $\,\chi\,$, the
 resulting quality factor is related to the lag via
 \ba
 Q^{-1}_{\textstyle{_{energy}}}~=~\sin|\delta|~~~
 \label{DI6_1}
 \label{DI6}
 \ea
 (not $\,\tan|\delta|~$ as commonly believed -- see the calculation in the Appendix to Efroimsky 2012).

 If however $\,E_{peak}(\chi)\,$ is introduced as the absolute maximum of $\,${\it{work}}$\,$ carried out on the sample at frequency
 $\,\chi\,$ over a time interval through which the power stays positive, then the appropriate $Q$ factor is connected to the lag via
 \ba
 Q^{-1}_{\textstyle{_{work}}}\;=\;\frac{\tan |\delta|}{1\;-\;\left(\;\frac{\textstyle \pi}{\textstyle 2}\;-\;
 |\delta|\;\right)  \;\tan|\delta|}\;\;\;,~~~~~
 \label{A81}
 \label{L14}
 \ea
 as was shown in {\it{Ibid}}.\footnote{In Efroimsky \& Williams (2009), $\,E_{peak}(\chi)\,$ was miscalled ``peak energy". However
 the calculation of $Q$ was performed there for $\,E_{peak}(\chi)\,$ introduced as the peak {\it{work}}.}

 The third definition of the quality factor (offered by Goldreich 1963) is
 \ba
 Q_{\textstyle{_{Goldreich}}}^{-1}=\,\tan|\delta |~~~.
 \label{Goldreich}
 \label{166}
 \ea
 This definition, though, corresponds neither to the peak work nor to the peak energy.

 In the limit of weak lagging, all three definitions entail
 \ba
 Q^{-1}\,=~
 |\delta|\;+\;O(\delta^2)\;\;\;.
 \label{DI6_2}
 \ea
 For the lag approaching $\,\pi/2\,$, the quality factor defined as (\ref{DI6_1}) assumes its minimal value,
 $\,Q_{\textstyle{_{energy}}}=1\,$, while definition (\ref{A81}) renders $\,Q_{\textstyle{_{work}}}=0\,$. The latter is natural,
 since in the considered limit the work performed on the system is negative, its absolute maximum being zero.\footnote{~As $\,Q<2\pi\,$
 implies $\,E_{peak}<\Delta E\,$, such small values of $\,Q\,$ are unattainable in the case of damped free oscillations.
 Still, $\,Q\,$ can assume such values under excitation, tides being the case.}

 In seismic studies or in exploration of attenuation in small samples, one's choice among the three definitions of $\,Q\,$ is a matter
 of personal taste, for the quality factor is large and the definitions virtually coincide.

 In the theory of tides, the situation is different, because at times one has to deal with situations where the definitions of $\,Q\,$
 disagree noticeably -- this happens when dissipation is intensive and $\,Q\,$ is of order unity. To make a choice, recall that the
 actual quantities entering the Fourier expansion of tides over the modes $\,\omega_{\textstyle{_{lmpq}}}\,$ are the
 products\footnote{~A historical tradition (originating from Kaula 1964) prescribes to denote the tidal phase lags with $\,\epsilon_{
 \textstyle{_{lmpq}}}\,$, while keeping for the dynamical Love numbers the same notation as for their static predecessors: $\,k_l\,$.
 These conventions are in conflict because the product $~k_l\,\sin\epsilon_{\textstyle{_{lmpq}}}~$ is the negative imaginary part of the
 complex Love number $\,\bar{k}_l\,$. More logical is to use the unified notation as in (\ref{25}). At the same time, it should not be
 forgotten that for triaxial bodies the functional form of the dependence of $\,\bar{k}_l\,$ on frequency is defined not only by $\,l\,$
 but also by $\,m,\,p,\,q\,$. In those situations, one has to deal with $\,k_{\textstyle{_{lmpq}}}\,\sin\epsilon_{\textstyle{_{lmpq}}}
 \,$, see Section \ref{ka}.}
 \ba
 k_l\,\sin\epsilon_{
 \textstyle{_{l}}}\,=\,k_{\textstyle{_l}}(\omega_{\textstyle{_{lmpq}}})\,\sin\epsilon_{\textstyle{_l}}(\omega_{\textstyle{_{lmpq}}})
 ~~~,
 \label{25}
 \ea
 where $\,k_{\textstyle{_l}}(\omega_{\textstyle{_{lmpq}}})\,$ are the dynamical analogues to the Love numbers. It is these products
 that show up in the $\,lmpq\,$ terms of the expansion for the tidal potential (force, torque). From this point of view, a definition
 like (\ref{DI6}) would be preferable, though this time with the tidal lag $\,\epsilon\,$ instead of the seismic lag $\,\delta~$:
 \begin{subequations}
 \ba
 Q^{-1}_{\textstyle{_{l}}}
 ~=~\sin|\,\epsilon_{\textstyle{_{l}}}
 \,|
 \label{pref_1}
 \ea
 or, in a more detailed manner:
 \ba
 Q^{-1}_{\textstyle{_l}}(\omega_{\textstyle{_{lmpq}}})~=~\sin|\,\epsilon_{\textstyle{_l}}(\omega_{\textstyle{_{lmpq}}})\,|~~~.
 \label{pref_2}
 \ea
 \label{pref}
 \end{subequations}
 Under this definition, one is free to substitute $\,k_l\,\sin\epsilon_{\textstyle{_{l}}}\,$ with $\,k_l/Q_{\textstyle{_{l}}}\,$. The
 subscript $\,l\,$ accompanying the tidal quality factor will then serve as a reminder of the distinction between the tidal
 quality factor and its seismic counterpart.

 While the notion of the tidal quality factor has some illustrative power and may be employed for rough estimates, calculations
 involving bodily tides should be based not on the knowledge of the quality factor but on the knowledge of the overall
 frequency dependence of products $\,k_l\,\sin\epsilon_{\textstyle{_{l}}}\,=\,k_{\textstyle{_l}}(\omega_{\textstyle{_{lmpq}}})\,\sin
 \epsilon_{\textstyle{_l}}(\omega_{\textstyle{_{lmpq}}})\,$. Relying on these functions would spare one of the ambiguity in definition
 of $\,Q\,$ and would also enable one to take into account the frequency dependence of the dynamical Love numbers.

 \section{The Andrade model and its reparameterisation}\label{Andrade_section}

 In the low-frequency limit, the mantle's behaviour is unlikely to differ much from that of the Maxwell body, because over timescales
 much longer than 1 yr viscosity dominates (Karato \& Spetzler 1990). At the same time, the accumulated geophysical, seismological,
 and geodetic observations suggest that at shorter timescales inelasticity takes over and the mantle is described by the Andrade model.
 However, the near-Maxwell behavior expected at low frequencies can be fit into the Andrade formalism, as we shall explain below.

 \subsection{Experimental data: the power scaling law}

 Dissipation in solids may be effectively modeled using the empirical scaling law
 \ba
 \sin\delta\;=\;\left(\,{\cal{E}}\,\chi\,\right)^{\textstyle{^{-p}}}~~,
 \label{generic}
 \ea
 $\cal{E}\,$ being a constant having the dimensions of time. This ``constant" may itself bear a (typically, much slower) dependence
 upon the frequency $\,\chi\,$. The dependence of $\,{\cal E}\,$ on the temperature is given by the Arrhenius law (Karato 2008).

 Experiments demonstrate that the power dependence (\ref{generic}) is surprisingly universal, with the exponential $\,p\,$ robustly
 taking values within the interval from $\,0.14\,$ to $\,0.4\,$ (more often, from $\,0.14\,$ to $\,0.3\,$).

 For the first time, dependence (\ref{generic}) was measured on metals in a lab. This was done by Andrade (1910), who also tried to pick up
 an expression for the compliance compatible with this scaling law. Later studies have demonstrated that this law works equally well,
 and with similar values of $\,p\,$, both for silicate rocks (Weertman \& Weertman 1975, Tan et al. 1997) and ices (Castillo-Rogez
 2009, McCarthy et al 2007).

 Independently from the studies of samples in the lab, the scaling behaviour (\ref{generic}) was obtained via measurements of
 dissipation of seismic waves in the Earth (Mitchell 1995, Stachnik et al. 2004, Shito et al. 2004).

 The third source of confirmation of the power scaling law came from geodetic experiments that included: (a) satellite laser ranging
 (SLR) of tidal variations in the $\,J_2\,$ component of the gravity field of the Earth; (b) space-based observations of tidal
 variations in the Earth's rotation rate; and (c) space-based measurements of the Chandler wobble period and damping (Benjamin et al.
 2006, Eanes \& Bettadpur 1996, Eanes 1995).\footnote{~It should be noted that in reality the geodetic measurements were confirming the
 power law (\ref{generic}) not for the seismic lag $\,\delta\,$ but for the tidal lag $\,\epsilon~$, an important detail to be addressed
 shortly.}

 While samples of most minerals furnish the values of $\,p\,$ lying within the interval $\,0.15\,-\,0.4\,$, the geodetic
 measurements give $\,0.14\,-\,0.2\,$. At least a fraction of this difference may be attributed to the presence of partial melt, which
 is known to have lower values of $\,p\,$ (Fontaine et al. 2005).

 On all these grounds, it is believed that mantles of terrestrial planets are adequately described by the Andrade model, at least in
 the higher frequency band where inelasticity dominates (Gribb \& Cooper 1998, Birger 2007, Efroimsky \& Lainey 2007, Zharkov \& Gudkova
 2009). Some of the other models were considered by Henning et al. (2009).

 The Andrade model is equally well applicable to celestial bodies with ice mantles (for application to Iapetus see Castillo-Rogez et al.
 2011) and to bodies with considerable hydration in a silicate mantle.\footnote{~Damping mechanisms in a wet planet will be the same as in a
 dry one, except that their efficiency will be increased. So the dissipation rate will have a similar frequency dependence but higher
 magnitude.} The model can also be employed for modeling of the tidal response {\it{of the solid parts}} of objects with significant
 liquid-water layers.{\footnote{~In the absence of internal oceans, a rough estimate of the tidal response can be obtained through
 modeling the body with a homogeneous sphere. However the presence of such oceans makes it absolutely necessary to calculate the
 overall response through integration over the solid and liquid layers. As demonstrated by Tyler (2009), tidal dissipation in
 internal ocean layers can play a big role in rotational dynamics of the body.}}

 \subsection{The Andrade model in the time domain}

 The compliance function of the Andrade body (Cottrell \& Aytekin 1947, Duval 1978),
 \ba
 J(t-t\,')\;=\;\left[\;J\;+\;(t-t\,')^{\alpha}~\beta+\;\left(t-t\,'\right)~{\eta}^{-1}\;\right]\,\Theta(t-t\,')~~~,
 \label{gh}
 \ea
 contains empirical parameters $\alpha\,$ and $\,\beta\,$, the steady-state viscosity $\eta\,$, and the unrelaxed compliance $\,J\equiv
 J(0)=1/\mu(0)\,=\,1/\mu\,$. We endow the right-hand side of (\ref{gh}) with the Heaviside step-function $\,\Theta(t-t\,')\,$, to
 ensure that insertion of (\ref{gh}) into (\ref{I12_4}), with the subsequent differentiation, yields the elastic term
 $\,J\,\delta(t-t\,')\,$ under the integral. The model allows for description of dissipation mechanisms over a continuum of frequencies,
 which is useful for complex materials with a range of grain sizes.

 The Andrade model can be thought of as the Maxwell model equipped with an extra term $\,(t-t\,')^{\alpha}\,\beta\,$ describing hereditary
 reaction of strain to stress. The Maxwell model
 \ba
 J^{\textstyle{^{(Maxwell)}}}(t-t\,')\;=\;\left[\;J\;+\;\left(t-t\,'
 \right)~{\eta}^{-1}\;\right]\,\Theta(t-t\,')~~~
 \label{Ma}
 \ea
 is a simple rheology, which has a long history of application to planetary problems, but generally has too strong a frequency
 dependence at higher frequencies where inelasticity becomes more efficient than viscosity (Karato 2008). Insertion of (\ref{Ma}) into
 (\ref{I12_4}) renders strain consisting of two separate inputs. The one proportional to $\,J\,$ implements the  instantaneous (elastic)
 reaction, while the one containing $\,{\eta}^{-1}\,$ is responsible for the viscous part of the reaction.

 Just as the viscous term $~\left(t-t\,'\right)\,{\eta}^{-1}~$ showing up in (\ref{gh} - \ref{Ma}), so the inelastic term
 $~(t-t\,')^{\alpha}\,\beta~$ emerging in the Andrade model (\ref{gh}) is delayed -- both terms reflect how the past stressing is
 influencing the present deformation. At the same time, inelastic reaction differs from viscosity both mathematically and physically,
 because it is produced by different physical mechanisms.

 A disadvantage of the formulation (\ref{gh}) of the Andrade model is that it contains a parameter of fractional dimensions, $\,\beta\,$.
 To avoid fractional dimensions, we shall express this parameter, following Efroimsky (2012), as
 \begin{subequations}
 \ba
 \beta\,=\,J~\tau_{_A}^{-\alpha}\,=~ \mu^{-1}\,\tau_{_A}^{-\alpha}~~~,
 \label{beta_01}
 \ea
 the new parameter $\,\tau_{_A}\,$ having dimensions of time. This is the timescale associated with the Andrade creep, wherefore it may be
 named as the ``Andrade time" or the ``inelastic time".

 Another option is to express $\,\beta\,$ as
 \ba
 \beta\,=\,{\zeta}^{-\alpha}\,J~\tau_{_M}^{-\alpha}\,=~\zeta^{-\alpha}\,\mu^{-1}\,\tau_{_M}^{-\alpha}~~~,
 \label{beta_02}
 \ea
 \label{beta_0}
 \end{subequations}
 where the dimensionless parameter $\,\zeta\,$ is related through
 \ba
 \zeta~=~\frac{\tau_{_A}}{\tau_{_M}}~~~
 \label{zeta}
 \ea
 to the Andrade timescale $\,\tau_{_A}\,$ and to the Maxwell time
 \ba
 \tau_{_M}\,\equiv\,\frac{\eta}{\mu}\,=\,\eta\,J~~.
 \label{}
 \ea
 In terms of the so-introduced parameters, the compliance assumes the form of
 \begin{subequations}
 \ba
 J(t-t\,')&=&J~\left[\;1\;+\;\left(\,\frac{t-t\,'}{\tau_{_A}}\,\right)^{\alpha}\,+~\frac{t-t\,'}{\tau_{_M}}\;\right]\,
 \Theta(t-t\,')\;\;\;
 \label{compliance_1}\\
 \nonumber\\
 \nonumber\\
 &=&J~\left[\;1\;+\;\left(\,\frac{t-t\,'}{\zeta~\tau_{_M}}\,\right)^{\alpha}\,+~\frac{t-t\,'}{\tau_{_M}}\;\right]\,
 \Theta(t-t\,')\;\;\;.
 \label{compliance_2}
 \ea
 \label{comliance}
 \label{I64}
 \end{subequations}
 For $\,\tau_{_A}\ll\,\tau_{_M}\,$ (or, equivalently, for $\,\zeta\,\ll\,1\,$), inelasticity plays a more important role than
 viscosity. On the other hand, a large $\,\tau_{_A}\,$ (or large $\,\zeta\,$) would imply suppression of inelastic creep, compared to
 viscosity.

 It has been demonstrated by Castillo-Rogez that under low stressing (i.e., when the grain-boundary diffusion is the
 dominant damping mechanism -- like in Iapetus) $\,\beta\,$ obeys the relation
 \begin{subequations}
 \ba
 \beta\,\approx~J~\tau_{_M}^{-\alpha}\,=\,J^{1-\alpha}\,\eta^{-\alpha}\,=\,\mu^{\alpha-1}
 \,\eta^{-\alpha}~~~,
 \label{beta_1}
 \ea
 (see, e.g., Castillo-Rogez et al. 2011). Comparing this to (\ref{beta_0}), we can say that the Andrade and Maxwell timescales are
 close to one another:
 \ba
 \tau_{_A}\,\approx~\tau_{_M}~~~
 \label{beta_2}
 \ea
 or, equivalently, that the dimensionless parameter $\,\zeta\,$ is close to unity:
 \ba
 \zeta~\approx~1~~~.
 \label{beta_3}
 \ea
 \label{beta}
 \end{subequations}
 Generally, we have no reason to expect the Andrade and Maxwell timescales to coincide, nor even to be comparable under all
 possible circumstances. While (\ref{beta}) may work when inelastic friction is determined mainly by the grain-boundary diffusion, we are also aware
 of a case when the timescales $\,\tau_{_A}\,$ and $\,\tau_{_M}\,$ differ considerably. This is a situation when stressing is stronger,
 and the inelastic part of dissipation is defined mainly by dislocations unpinning (i.e., by the Andrade creep). This is what happens in mantles of earths and superearths.

 On theoretical grounds, Karato \& Spetzler (1990) point out that the dislocation-unpinning mechanism remains effective in the Earth's
 mantle down to the frequency threshold $\,\chi_{\textstyle{_0}}\sim 1~\mbox{yr}^{-1}$. At lower frequencies, this mechanism becomes less efficient,
 giving way to viscosity. Thus at low frequencies the mantle's behaviour becomes closer to that of the Maxwell body.\footnote{~Using the
 Andrade model as a fit to the experimentally observed scaling law (\ref{generic}), we see that the exponential $\,p\,$ coincides with the
 Andrade parameter $\,\alpha<1\,$ at frequencies above the said threshold, and that $\,p\,$ becomes closer to unity below the
 threshold -- see subsection \ref{ggg} below.} This important example tells us that the Andrade time $\,\tau_{_A}\,$ and the
 dimensionless parameter $\,\zeta\,$ may, at times, be more sensitive to the frequency than the Maxwell time would be. Whether $\,\tau_{_A}
 \,$ and $\,\zeta\,$ demonstrate this sensitivity or not -- may, in its  turn, depend upon the intensity of loading, i.e., upon the damping
 mechanisms involved.

 \subsection{The Andrade model in the frequency domain}\label{o}

 Through (\ref{inverse}), it can be demonstrated (Findley et al. 1976) that in the frequency domain the compliance of an Andrade material
 reads as
 \begin{subequations}
 \ba
 {\bar{\mathit{J\,}}}(\chi)&=&J\,+\,\beta\,(i\chi)^{-\alpha}\;\Gamma\,(1+\alpha)\,-\,\frac{i}{\eta\chi}
 \label{112_1}\\
 \nonumber\\
 &=&J\,\left[\,1\,+\,(i\,\chi\,\tau_{_A})^{-\alpha}\;\Gamma\,(1+\alpha)~-~i~(\chi\,\tau_{_M})^{-1}
 \right]\;\;\;,
 \label{112_2}\\
 \nonumber\\
 &=&J\,\left[\,1\,+\,(i\,\chi\,\zeta\,\tau_{_M})^{-\alpha}\;\Gamma\,(1+\alpha)~-~i~(\chi\,\tau_{_M})^{-1}
 \right]\;\;\;,
 \label{112_3}
 \ea
 \label{112}
 \label{LL44}
 \end{subequations}
 $\chi\,$ being the frequency, and $\,\Gamma\,$ denoting the Gamma function. The imaginary and real parts of the complex
 compliance are:
 \begin{subequations}
 \ba
 {\cal I}{\it m} [ \bar{J}(\chi)]&=&-\;\frac{1}{\eta\,\chi}\;-\;\chi^{-\alpha}\,\beta\;\sin\left(
 \,\frac{\alpha\,\pi}{2}\,\right)
 \;\Gamma(\alpha\,+\,1)
 ~~~~~~~~~~~~~~~\quad\quad\quad
 \label{A3a}\\
 \nonumber\\
 &=&-\;J\,(\chi\,\tau_{_M})^{-1}\;-\;J\,(\chi\,\tau_{_A})^{-\alpha}\;\sin\left(
 \,\frac{\alpha\,\pi}{2}\,\right)\;\Gamma(\alpha\,+\,1)~~~~~~~~~~~~~~~\quad\quad\quad
 \label{A3b}\\
 \nonumber\\
 &=&-\;J\,(\chi\,\tau_{_M})^{-1}\;-\;J\,(\chi\,\zeta\,\tau_{_M})^{-\alpha}\;\sin\left(
 \,\frac{\alpha\,\pi}{2}\,\right)\;\Gamma(\alpha\,+\,1)~~~~~~~~~~~~~~~\quad\quad\quad
 \label{A3c}
 \ea
 \label{A3}
 \end{subequations}
 and
 \begin{subequations}
 \ba
 {\cal R}{\it e} [ \bar{J}(\chi)]&=&J\;+\;\chi^{-\alpha}\,\beta\;\cos\left(\,\frac{\alpha\,\pi}{2}
 \,\right)\;\Gamma(\alpha\,+\,1)~~~~~~~~~~~~~~~~~~~~~~~~~~~~~~
 \label{A4a}\\
 \nonumber\\
 &=&J\;+\;J\,(\chi\tau_{_A})^{-\alpha}\;\cos\left(\,\frac{\alpha\,\pi}{2}\,\right)
 \;\Gamma(\alpha\,+\,1)~~~\quad\quad\quad\quad\quad~\quad\quad\quad\quad\quad\quad
 \label{A4b}\\
 \nonumber\\
 &=&J\;+\;J\,(\chi\,\zeta\,\tau_{_M})^{-\alpha}\;\cos\left(\,\frac{\alpha\,\pi}{2}\,\right)
 \;\Gamma(\alpha\,+\,1)~~~.\quad\quad\quad\quad\quad~\quad\quad\quad\quad\quad\quad
 \label{A4c}
 \ea
 \label{A4}
 \end{subequations}
 The ensuing frequency dependence of the phase lag will look as$\,$\footnote{~In some publications (e.g., Nimmo 2008), formula
 (\ref{113a}) is given as an expression for the inverse quality factor. This is legitimate when the latter is defined through (\ref{Goldreich}).}
 \begin{subequations}
 \ba
 \tan\delta(\chi)&=&-~\frac{{{\cal{I}}\textit{m}\left[\bar{J}(\chi)\right]}}{{\cal{R}}{\textit{e}\left[\bar{J}(\chi)\right]}}
 ~=~\frac{(\eta\;\chi)^{\textstyle{^{\,-1}}}\,+~\chi^{\textstyle{^{\,-\alpha}}}\;\beta\;\sin\left(\frac{
 \textstyle\alpha~\pi}{\textstyle 2}\right)\Gamma\left(\alpha\,+\,1\right)}{\mu^{\textstyle{^{~-1}}
 }\,+\;\chi^{\textstyle{^{\,-\alpha}}}\;\beta\;\cos\left(\frac{\textstyle\alpha\;\pi}{\textstyle 2}
 \right)\;\Gamma\left(\textstyle\alpha\,+\,1\right)}
 \label{113a}
 \label{LL45a}
  ~\\ \nonumber\\ \nonumber\\
 &=&\frac{ \frac{\textstyle 1}{\textstyle \,\chi~\tau_{_M}\,}\,+\,\left(\frac{\textstyle 1}{\textstyle \,\chi~\tau_{_A}\,}\right)^{
 \alpha}\,~\Gamma\left(\alpha\,+\,1\right)~\,\sin\left(\frac{\textstyle\alpha~\pi}{\textstyle 2}\right)}{1\,+~\left(\,\frac{\textstyle
 1}{\textstyle \,\chi~\tau_{_A}\,}\,\right)^{\alpha}~\,\Gamma\left(\textstyle\alpha\,+\,1\right)
 \,~\cos\left(\frac{\textstyle\alpha\;\pi}{\textstyle 2}\right)}~
 \label{Andra}
 =~\frac{z^{-1}\,\zeta\,+\,z^{-\alpha}\,\sin\left(\frac{
 \textstyle\alpha~\pi}{\textstyle 2}\right)\Gamma\left(\alpha\,+\,1\right)}{1\,+~z^{-\alpha}
 \,\cos\left(\frac{\textstyle\alpha\;\pi}{\textstyle 2}\right)~\Gamma\left(\textstyle\alpha\,+\,1\right)}~~~,\quad\quad\quad
 \label{113b}
 \label{LL45b}
 \ea
 \label{113}
 \label{LL45}
 \end{subequations}
 with $z\,$ being the dimensionless frequency defined through
 \ba
 z~\equiv~\chi~\tau_{_A}~=~\chi~\tau_{_M}~\zeta~~~.
 \label{sim}
 \ea
 Evidently, for $\,\beta\rightarrow 0\,$ (that is, for $\,\zeta\rightarrow\infty\,$ or $\,\tau_{_A}\rightarrow
 \infty\,$), expression (\ref{LL45}) approaches
 \ba
 \tan\delta(\chi)\,=\,\left(\tau_{_M}\,\chi\right)^{\textstyle{^{\,-1}}}~~~,
 \label{MMa}
 \ea
 which is the frequency dependence appropriate to the Maxwell body.

 \subsection{{Low frequencies: from Andrade toward Maxwell}}{\label{ggg}}

 The Andrade body demonstrates the so-called ``elbow dependence" of the dissipation rate upon frequency. At high frequencies, the lag
 $\,\delta\,$ satisfies the power law
 \ba
 \tan\delta~\sim~\chi^{-p}~~~,
 \label{}
 \ea
 the exponential being expressed via an empirical parameter $\,\alpha\,$, where $\,0\,<\,\alpha\,<\,1\,$ for most materials. It follows
 from (\ref{LL45}) that at higher frequencies $\,p=\alpha\,$, while at low frequencies
 $\,p={1-\alpha}\,$.

 The statement by Karato \& Spetzler (1990), that the mantle's behaviour at low frequencies should lean toward that of the Maxwell
 body, can be fit into the Andrade formalism, if we agree that at low frequencies either $\,\alpha\,$ approaches zero (so $p$
 approaches unity) or $\,\zeta\,$ becomes large (so $\,\tau_{_A}\,$ becomes much larger than $\,\tau_{_M}\,$). The latter option is
 more physical, because the increase of $\,\tau_{_A}\,$ would reflect the slowing-down of the unpinning mechanism studied in
 {\it{Ibid}}.

 One way or another, the so-parameterised Andrade model is fit to embrace the result from {\it{Ibid}}. Hence our treatment will permit
 us to describe both the high-frequency range where the traditional Andrade model is applicable, and the low-frequency band where the
 behaviour of the mantle deviates from the Andrade model toward the Maxwell body. Comparison of the two models in the frequency domain
 is presented in Figure 1.


 \begin{figure}[h]
  {\includegraphics[width=17.1cm]{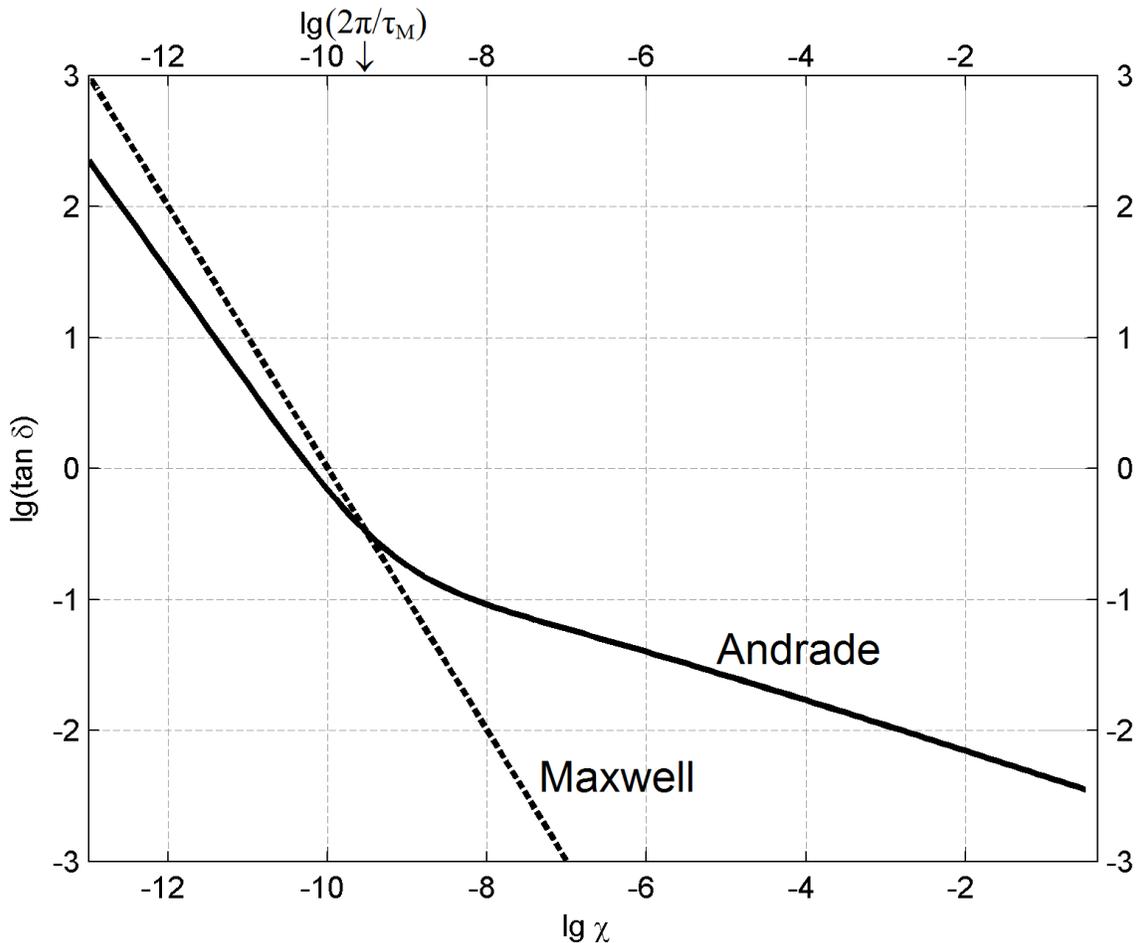}}
 {\caption{\small{Andrade and Maxwell models in the frequency domain. The plot shows the decadic logarithm of $\,\tan\delta\,$, as a
 function of the decadic logarithm of the forcing frequency $\,\chi\,$ (in cycles s$^{-1}$). For the Andrade body, the tangent of phase lag is given by
 (\ref{Andra}), with $\,\alpha=0.2\,$ and $\,\tau_{_A}=\tau_{_M}=10^{10}\,$s. For the Maxwell body, the tangent is rendered by
 (\ref{MMa}), with $\,\tau_{_M}=10^{10}\,$s.}}}
 \label{Figure_1}
 \end{figure}
 \hspace{1.3cm}



 \section{Expanding a tidal potential or torque -- over the tidal modes or over the forcing frequencies?}\label{ka}

 Consider a binary system with mean motion $\,n\,$, and suppose that tidal dissipation in one of the bodies much exceeds that in its
 companion. Then the former body may be treated as the tidally perturbed primary, the latter being its tide-raising secondary. The
 sidereal angle and the spin rate of the primary will be expressed with $\,\theta\,$ and $\,\stackrel{\bf\centerdot}{\theta\,}\,$,
 while the node, pericentre, and mean anomaly of the secondary, as seen from the primary,\footnote{~When the role of the primary
 is played by a planet and the role of the perturbing secondary is played by the host star,
 the argument of the pericentre of the star as seen from the planet, $\,\omega\,$, differs by $\,\pi\,$ from the argument of the
 pericentre of the planet as seen from the star.

 Also mind that in equation (\ref{C}) the letter $\,\omega\,$ with the subscript $\,lmpq\,$ denotes, as ever, a tidal mode, while the
 same letter without a subscript stands for the periapse. The latter use of this letter in equation (\ref{C}) is exceptional, in that
 elsewhere in the paper the letter $\,\omega\,$, with or without a subscript, always denotes a tidal mode.} will be denoted by $\,\Omega
 \,$, $\omega\,$, and ${\cal M}$.

 In the Darwin-Kaula theory, bodily tides are expanded over the modes
 \ba
 \omega_{lmpq}\;\equiv\;(l-2p)\;\dot{\omega}\,+\,(l-2p+q)\;\dot{\cal{M}}\,+\,m\;(\dot{\Omega}\,-\,\dot{\theta})\,\approx\,
 (l-2p+q)\;n\,-\,m\;\dot{\theta}
 ~~~,~~~
 \label{C}
 \ea
 with $l,\,m,\,p,\,q\,$ being integers. Dependent upon the values of the mean motion, spin rate, and the indices, the tidal modes
 $\,\omega_{{\it l}mpq}\,$ may be positive or negative or zero.

 In the expansion of the tidal potential or torque or force, summation over the integer indices goes as $~\sum_{l=2}^{\infty}\sum_{m=0}^{l}
 \sum_{p=0}^{\infty}\sum_{q=\,-\,\infty}^{\infty}~.$ ~For example, the secular polar component of the tidal torque will read as:
 \ba
 {\cal T}~=~\sum_{l=2}^{\infty}\sum_{m=0}^{l}\sum_{p=0}^{\infty}\sum_{q=\,-\,\infty}^{\infty}.~.~.~\,
 k_{\textstyle{_{l}}}(\omega_{\textstyle{_{lmpq}}})~\sin\epsilon_{\textstyle{_{l}}}(\omega_{\textstyle{_{lmpq}}})~~~,
 \label{torque}
 \ea
 where the ellipsis denotes a function of the primary's radius and the secondary's orbital elements. The functions $\,
 k_{\textstyle{_{l}}}(\omega_{\textstyle{_{lmpq}}})\,$ are the dynamical analogues of the static Love numbers, while the
 phase lags corresponding to the tidal modes $\,\omega_{\textstyle{_{lmpq}}}\,$ are given by
 \ba
 \epsilon_{\textstyle{_{l}}}(\omega_{\textstyle{_{lmpq}}})
 ~=~\omega_{\textstyle{_{lmpq}}}~\Delta t_{\textstyle{_l}}(|\omega_{\textstyle{_{lmpq}}}|)~=~
 |\,\omega_{\textstyle{_{lmpq}}}\,|~\Delta t_{\textstyle{_l}} (|\omega_{\textstyle{_{lmpq}}}|)~\mbox{sgn}\,\omega_{\textstyle{_{lmpq}}}~
 =~\chi_{\textstyle{_{lmpq}}}~\Delta t_{\textstyle{_l}} (\chi_{\textstyle{_{lmpq}}})~\,\mbox{sgn}\,\omega_{\textstyle{_{lmpq}}}~~~.~~~~
 \label{phaselag}
 \ea
 Here the positively defined quantities
 \ba
 \chi_{\textstyle{_{lmpq}}}\,\equiv~|\,\omega_{\textstyle{_{lmpq}}}\,|~~~
 \label{B}
 \ea
 are the forcing frequencies in the material, while the positively defined time lags $\,\Delta t_{\textstyle{_{lmpq}}}\,$ are their
 functions.

 Following Kaula (1964), the phase and time lags are often denoted with $\,\epsilon_{lmpq}\,$ and $\,\Delta t_{lmpq}\,$. For
 near-spherical bodies, though, the notations $~\epsilon_{{\it l}}(\chi_{{\it l}mpq})~$ and $\,\Delta t_{{\it l}}(\chi_{{\it l}mpq})\,$
 would be preferable, because for such bodies the functional form of the dependency $\,\epsilon_{lmpq}(\chi)\,$ is defined by
 {\it{l}}{\small{$\,$s}}, but is ignorant of the values of the other three indices.\footnote{~Within the applicability realm of the
 elastic-viscoelastic analogy employed in subsection \ref{pri} below, the functional form of the complex Love number $\,\bar{k}_l(\chi)
 \,$ of a near-spherical object is determined by index $\,\it l\,$ solely, while the integers $\,m,\,p,\,q\,$ show up through the value
 of the frequency: $~\bar{k}_{{\it l}}(\chi)\,=\,\bar{k}_{{\it l}}(\chi_{{\it l}mpq})~$. This applies to the
 lag too, since the latter is related to $\,\bar{k}_{\it l}\,$ via (\ref{ggffrr}).

 For triaxial bodies, the functional forms of the frequency dependencies of the Love numbers and phase lags do depend upon $\,m,\,p,\,q\,$,
 because of coupling between spherical harmonics. In those situations, notations $\,\bar{k}_{{\it l}mpq}\,$ and $\,\epsilon_{{\it l}mpq}\,$
 become necessary (Dehant 1987a,b; $\,$Smith 1974). The Love numbers of a slightly non-spherical primary differ from the Love numbers of
 the spherical reference body by a term of the order of the flattening, so a small non-sphericity can usually be ignored.\label{HH}}
 The same therefore applies to the time lag.

 The forcing frequencies in the material of the primary, $\,\chi_{\textstyle{_{lmpq}}}\,$, are positively defined. While the general
 formula for a Fourier expansion of a field includes integration (or summation) over both positive and negative frequencies, it is easy
 to demonstrate that in the case of real fields it is sufficient to expand over positive frequencies only. The condition of the field
 being real requires that the real part of a Fourier term at a negative frequency is equal to the real part of the term at an opposite,
 positive frequency. Hence one can get rid of the terms with negative frequencies, at the cost of doubling the appropriate terms with
 positive frequencies. (The convention is that the field is the real part of a complex expression.)

 The tidal theory is a rare exception to this rule: here, a contribution of a Fourier mode into the potential is not completely
 equivalent to the contribution of the mode of an opposite sign. The reason for this is that the tidal theory is developed to render
 expressions for tidal forces and torques, and the sign of the tidal mode $\,\omega_{\textstyle{_{lmpq}}}\,$ shows up explicitly in
 those expression. This happens because the phase lag in (\ref{torque}) is the product (\ref{phaselag}) of the tidal mode $\,
 \omega_{\textstyle{_{lmpq}}}\,$ and the positively defined time lag $\,\Delta t_{\textstyle{_{lmpq}}}\,$.

 This way, if we choose to expand tide over the positively defined frequencies $\,\chi\,$ only, we shall have to insert ``by hand" the
 multipliers
 \ba
 \mbox{sgn}\,\omega_{\textstyle{_{lmpq}}}~=~\mbox{sgn}\left[~({\it l}-2p+q)\,n~-~m\,\dot{\theta}~\right]
 \label{multi}
 \ea
 into the expressions for the tidal torque and force, a result to be employed below in formula (\ref{L39bbb}).
 The topic is explained in greater detail in Efroimsky (2012).

 \section{Complex Love numbers and the elastic-viscoelastic analogy}\label{compla}

 Let us recall briefly the switch from the stationary Love numbers to their dynamical counterparts, the Love operators. The method was
 pioneered, probably, by Zahn (1966) who applied it to a purely viscous planet. The method works likewise for an arbitrary linear
 rheological model, insofar as the elastic-viscoelastic analogy (also referred to as the correspondence principle) remains in force.

  \subsection{From the Love numbers to the Love operators}\label{3.3}\label{4.1}

 A homogeneous near-spherical incompressible primary alters its shape and potential, when influenced by a static point-like secondary.  At
 a point $\Rbold = (R,\lambda,\phi)$, the potential due to a tide-generating secondary of mass $M^*_{sec}\;$, located at $\,{\erbold}^{~*}=
 (r^*,\,\lambda^*,\,\phi^*)\,$ with $\,r^*\geq R\,$, can be expressed through the Legendre polynomials $\,P_{\it l}(\cos\gamma)\,$ or the
 Legendre functions $\,P_{{\it{l}}m}(\sin\phi)\,$:
 \ba
 \nonumber
 W(\eRbold\,,\,\erbold^{~*})&=&\sum_{{\it{l}}=2}^{\infty}~W_{\it{l}}(\eRbold\,,~\erbold^{~*})~=~-~\frac{G\;M^*_{sec}}{r^{
 \,*}}~\sum_{{\it{l}}=2}^{\infty}\,\left(\,\frac{R}{r^{~*}}\,\right)^{\textstyle{^{\it{l}}}}\,P_{\it{l}}(\cos \gamma)~~~~\\
 \nonumber\\
 &=&-\,\frac{G~M^*_{sec}}{r^{\,*}}\sum_{{\it{l}}=2}^{\infty}\left(\frac{R}{r^{~*}}\right)^{\textstyle{^{\it{l}}}}\sum_{m=0}^{\it l}
 \frac{({\it l}-m)!}{({\it l}+m)!}(2-\delta_{0m})P_{{\it{l}}m}(\sin\phi)P_{{\it{l}}m}(\sin\phi^*)~\cos m(\lambda-\lambda^*)~~,\,\quad~~
 \quad
 \label{1}
 \label{L1}
 \ea
 $G\,$ being Newton's gravitational constant, and $\gamma\,$ being the angle between the vectors $\,{\erbold}^{\;*}\,$ and $\,\Rbold\,$
 originating from the primary's centre. The latitudes $\phi,\,\phi^*$ are reckoned from the primary's equator, while the longitudes
 $\lambda,\,\lambda^*$ are reckoned from a fixed meridian on the primary.

 The $\,{\it{l}}^{th}$ spherical harmonic $\,U_l(\erbold)\,$ of the resulting change of the primary's potential at an exterior point
 $\,\erbold\,$ is connected to the $\,{\it{l}}^{th}$ spherical harmonic $\,W_l(\Rbold\,,\,\erbold^{\;*})\,$ of the perturbing exterior
 potential via $~U_{\it l}(\erbold)=\left({R}/{r}\right)^{{\it l}+1}{k}_{\it l}\,W_{\it{l}}(\eRbold\,,\,\erbold^{\;*})~$, so the total
 change in the exterior potential of the primary becomes:
  \ba
 U(\erbold)~=~\sum_{{\it l}=2}^{\infty}~U_{\it{l}}(\erbold)~=~\sum_{{\it l}=2}^{\infty}\,\left(\,\frac{R}{r}\,\right)^{{\it l}+1}
 k_{l}\;\,W_{\it{l}}(\eRbold\,,\;\erbold^{\;*})~~~.~~~~~~~
 ~~~~~~~~~~~~~~~~
 \label{2}
 \label{L2}
 \ea
  While in (\ref{L1})
 $\,\eRbold
 \,$ could lie either outside or on the surface of the primary, in (\ref{L2}) it would be both convenient and conventional to make
 $\,\eRbold\,$ a surface point. Both in (\ref{L1}) and (\ref{L2}), the vector $\,\erbold
 \,$ denotes an exterior point located above the surface point $\,\eRbold\,$ at a radius $\,r\,\geq\,R\,$ (with the same latitude and
 longitude), while $\,\erbold^{\;*}
 \,$ signifies the position of the tide-raising secondary. The quantities $\,k_{l}\,$ are the static Love numbers.

 Under dynamical stressing, the Love numbers turn into operators:
 \ba
 \label{L29}
 U_{\it l}(\erbold,\,t)\;=\;\left(\,\frac{R}{r}\,
 \right)^{{\it l}+1}\hat{k}_{\it l}(t)\;W_{\it{l}}(\eRbold\,,\;\erbold^{\;*},\,t\,')~~~,
 \ea
 where integration over the semi-interval $\,t\,'\in\left(\right.-\infty,\,t\,\left.\right]\,$ is implied:
 \begin{subequations}
 \label{subequations_3}
 \label{L30}
 \ba
 U_{\it l}(\erbold,\,t)&=&\left(\frac{R}{r}
 \right)^{{\it l}+1}\int_{t\,'=-\infty}^{t\,'=t} k_{\it l}(t-t\,')\stackrel{\bf \centerdot}{W}_{\it{l}}
 (\eRbold\,,\,\erbold^{\;*},\,t\,')\,dt\,'~~~~
 \label{churchur}
 \label{L30a}\\
 \nonumber\\
 \nonumber\\
 &=&\left(\frac{R}{r}
 \right)^{{\it l}+1}\,\left[k_l(0)W_l(t)\,-\,k_l(\infty)W_l(-\infty)\right]\,+\,\left(\frac{R}{r}
 \right)^{{\it l}+1}\int_{-\infty}^{t}
 {\bf\dot{\it{k}}}_{\textstyle{_l}}(t-t\,')\,~W_{\it{l}}
 (\eRbold\,,\,\erbold^{\;*},\,t\,')\,dt\,'~~.\quad\quad\quad
 \label{}
 \ea
 Like in the compliance operator (\ref{L18} - \ref{I12_3}), here we too obtain the boundary terms: one corresponding to the instantaneous
 elastic reaction, $~k_l(0)W(t)~$, another caused by the perturbation in the infinite past, $~-k_l(\infty)W(-\infty)~$. The latter term
 can be dropped by setting $\,W(-\infty)\,$ zero, while the former term may be included into the kernel in the same manner as in
 (\ref{conv} - \ref{I12_4}):
 \ba
 \nonumber
 \left(\frac{R}{r}
 \right)^{{\it l}+1}\,k_l(0)W_l(t)\,+\,\left(\frac{R}{r}
 \right)^{{\it l}+1}\int_{-\infty}^{t}
 {\bf\dot{\it{k}}}_{\textstyle{_l}}(t-t\,')\,~W_{\it{l}}
 (\eRbold\,,\,\erbold^{\;*},\,t\,')\,dt\,'\quad\quad\quad\quad\\
 \nonumber\\
 =\left(\frac{R}{r}\right)^{{\it l}+1}\int_{-\infty}^{t} \frac{d}{dt}\left[~{k}_{\it l}(t\,-\,t\,')~-~{k}_{\it l}(0)~+~
 {k}_{\it l}(0)\Theta(t\,-\,t\,')~\right]~W_{\it{l}}(\eRbold,\erbold^{\;*},t\,')\,dt\,'\quad.\quad\quad\quad\quad
 \label{churban}
 \ea
 \label{L30b}
 \end{subequations}
 All in all, neglecting the unphysical term with $\,W_l(-\infty)~$, and inserting the elastic term into the Love number
 not as $\,k_{l}(0)\,$ but as $\,{k}_{l}(0)\,\Theta(t-t\,')\,$, we arrive at
 \ba
 U_{\it l}(\erbold,\,t)\;=\;\left(\frac{R}{r}
 \right)^{{\it l}+1}\int_{-\infty}^{t} {\bf\dot{\it{k}}}_{\textstyle{_l}}(t-t\,')~W_{\it{l}}
 (\eRbold\,,\;\erbold^{\;*},\;t\,')\,dt\,'~,
 \label{chuk}
 \ea
 with $\,{k}_{\it l}(t-t\,')\,$ now incorporating the elastic reaction as $\,{k}_{l}(0)\,\Theta(t-t\,')\,$ instead of $\,{k}_{l}(0)\,$.
 For a perfectly elastic primary, the elastic reaction would be the only term present in the expression for $\,{k}_{\it l}(t-t\,')\,$.
 Then the time derivative of $\,{k}_{l}\,$ would be $\,{\bf{\dot{\it{k}}}}_{\textstyle{_{\it{l}}}}(t-t\,')\,=\,k_{\it l}\,\delta(t-
 t\,')\,$, with $\,k_{\it l}\,\equiv\,k_{\it l}(0)\,$ being the static Love number, and (\ref{chuk}) would reduce to $~U_{\it l}(\erbold
 ,\,t)\;=\;\left(\frac{\textstyle R}{\textstyle r}\right)^{{\it l}+1}k_l\,W_{\it{l}}(\eRbold\,,\;\erbold^{\;*},\;t)~$, like in the
 static case.

 Similarly to (\ref{L20}), the complex Love numbers are defined as the Fourier images of
 $\,\stackrel{\bf\centerdot}{k}_{\textstyle{_l}}(\tau)~$:
 \ba
 \label{L31}
 \int_{0}^{\infty}\bar{k}_{\textstyle{_l}}(\chi)e^{\inc\chi \tau}d\chi\;=\;
 \stackrel{\bf\centerdot}{
 k}_{\textstyle{_l}}(\tau)
 ~~~,
 \ea
 overdot denoting $\,d/d\tau\,$. Following Churkin (1998), the time derivatives $\,\stackrel{\bf\centerdot}{k}_{\textstyle{_{\,l}}}(t)\,$
 can be named {\emph{Love functions}}.\footnote{~Churkin (1998) used functions, which he denoted $\,k_{\it l}(t)\,$ and which were, due to
 a difference in notations, the same as our $\,\stackrel{\bf\centerdot}{k}_{\textstyle{_l}}(\tau)\,$.} Inversion of (\ref{L31}) renders:
 \ba
 \bar{k}_{\textstyle{_{l}}}(\chi)~=~\int_{0}^{\infty}
 {\bf\dot{\mbox{\it{k}}}}_{\textstyle{_{l}}}(\tau)\;\,e^{-\inc\chi \tau}\,d\tau~=~k_{\textstyle{_{l}}}(0)\;+\;
 \inc~\chi~\int_{0}^{\infty}\left[\,k_{\textstyle{_{l}}}(\tau)\,-\,k_{\textstyle{_{l}}}(0)\,\Theta(\tau)\,\right]\;
                e^{-\inc\chi \tau}\,d\tau~~~,~~~~
 \label{L32}
 \label{gig}
 \ea
 where integration from $\;{0}\,$ is sufficient, as the future disturbance contributes nothing to the present distortion, wherefore $\,k_{
 \it l}(\tau)\,$ vanishes at $\,\tau<0\,$. Recall that the time $\,\tau\,$ denotes the difference $t-t\,'$ and thus is reckoned from the
 present moment $t$ backward into the past

 In the frequency domain, (\ref{subequations_3}) will take the shape of
 \ba
 \bar{U}_{\textstyle{_{l}}}(\chi)\;=\;\left(\,\frac{R}{r}\,\right)^{l+1}\bar{k}_{\textstyle{_{l}}}(\chi)\;\,\bar{W}_{\textstyle{_{l}}}(\chi)\;\;\;,
 \label{VR}
 \label{L33}
 \ea
 $\chi\,$ being the frequency, while $\,\bar{U}_{\textstyle{_{l}}}(\chi)\,$ and $\,\bar{W}_{\textstyle{_{l}}}(\chi)\,$ being the Fourier or
 Laplace components of the potentials $\,{U}_{\textstyle{_{l}}}(t)\,$ and $\,{W}_{\textstyle{_{l}}}(t)\,$. The frequency-dependencies
 $\,\bar{k}_{\textstyle{_{l}}}(\chi)\,$ should be derived from the expression for $\,\bar{J}(\chi)\,$ or $\,\bar{\mu}(\chi)=1/\bar{J}(\chi)
 \,$. These expressions follow from the rheological model of the medium.

 Rigorously speaking, we ought to assume in expressions (\ref{L31} - \ref{L33}) that the spectral components are functions of the tidal
 mode $\,\omega\,$ and not of the forcing frequency $\,\chi\,$. However, as explained in the end of Section \ref{ka}, employment of
 the positively defined forcing frequencies is legitimate, insofar as we do not forget to attach the sign multipliers (\ref{multi}) to
 the terms of the Darwin-Kaula expansion for the tidal torque. Therefore here and hereafter we shall expand over $\,\chi\,$, with the said
 caveat kept in mind.

 \subsection{Complex Love numbers as functions of the complex compliance. The elastic-viscoelastic analogy}\label{pri}

 The dependence of the static Love numbers on the static rigidity modulus $\,\mu(\infty)
 \,$ looks as
 \ba
 k^{\textstyle{^{(static)}}}_{\it l}\,=\;\frac{3}{2\,({\it l}\,-\,1)}\;\,\frac{1}{1\;+\;A^{\textstyle{^{(static)}}}_{\it l}}~~~,
 \label{A_def}
 \ea
 where
 \ba
 A^{\textstyle{^{(static)}}}_{\it l}\,\equiv~\frac{\textstyle{(2\,{\it{l}}^{\,2}\,+\,4\,{\it{l}}\,+\,3)}}{\textstyle{{\it{l}}\,\mbox{g}\,
 \rho\,R}}~\,\mu(\infty)
 ~=\;\frac{\textstyle{3\;(2\,{\it{l}}^{\,2}\,+\,4\,{\it{l}}\,+\,3)}}{\textstyle{4\;{\it{l}}\,\pi\,
 G\,\rho^2\,R^2}}~\,\mu(\infty)~
 =~\frac{\textstyle{3\;(2\,{\it{l}}^{\,2}\,+\,4\,{\it{l}}\,+\,3)}}{\textstyle{4\;{\it{l}}\,
 \pi\,G\,\rho^2\,R^2~\,J(\infty)
 }~}~\,~~~,\quad~~
 \label{L4}
 \ea
 with $\,\rho\,$, g, and $R$ being the density, surface gravity, and radius of the body, and $G$ being the Newton gravitational constant. The
 static rigidity modulus and its inverse, the static compliance, are denoted here with $\,\mu(\infty)\,$ and $\,J(\infty)\,$,
 correspondingly. These notations imply that we identify {\it{static}} with {\it{relaxed}}.

 Specifically, the static quadrupole Love number will read:
 \ba
 k^{\textstyle{^{(static)}}}_{2}\,=\;\frac{3}{2}\;\,\frac{1}{1\;+\;A^{\textstyle{^{(static)}}}_{2}}~~~,
 \label{Mga}
 \ea
 where the quantity
 \ba
 A^{\textstyle{^{(static)}}}_2=\,\frac{\textstyle 57}{\textstyle 8\,\pi}~\frac{\textstyle \mu(\infty)}{\textstyle
 G\,\rho^2\,R^2}
 \label{A2}
 \ea
 is sometimes referred to as $\,\tilde{\mu}\,$. Clearly, $\,A^{\textstyle{^{(static)}}}_2\,$
 in (\ref{Mga}), as well as $\,A^{\textstyle{^{(static)}}}_l\,$
 in (\ref{A_def}), is a dimensionless measure of strength-dominated versus gravity-dominated
 behaviour.

 It is not immediately clear whether the same expression interconnects also $\,\bar{k}_{\it l}(\chi)\,$ with $\,\bar{\mu}(\chi)\,$.
 Fortunately, though, a wonderful theorem called {\it{elastic-viscoelastic analogy}}, also known as the {\it{correspondence principle}},
 ensures that the viscoelastic operational moduli $\,\bar{\mu}(\chi)\,$ or $\,\bar{J}(\chi)\,$ obey the same algebraic relations as the
 elastic parameters $\,\mu\,$ or $\,J\,$ ~(see, e.g., Efroimsky 2012 and references therein). For this reason, the Fourier or Laplace
 images of the viscoelastic equation of motion~\footnote{~In the equation of motion, we should neglect the acceleration term and the
 nonconservative inertial forces. Both neglects are justified at realistic frequencies (for details see the Appendix to Efroimsky 2012).}
 and of the constitutive equation look as their static counterparts, except that the stress, strain, and potentials get replaced with
 their Fourier or Laplace images, while $\,k_l\,$, $\,\mu\,$, and $\,J\,$ get replaced with the Fourier or Laplace images of $\,
 \stackrel{\bf\centerdot}{k}_{\textstyle{_l}}(t-t\,')~$, $\,\stackrel{\,\bf\centerdot}{\mu}(t-t\,')~$, and $\,\stackrel{~\bf\centerdot}
 {J}(t-t\,')~$. For example, the constitutive equation will look like: $~\bar{\sigma}_{\textstyle{_{\gamma\nu}}}~=~2~\bar{\mu}~\bar{u}
 _{\textstyle{_{\gamma\nu}}}~$. Therefore the solution to the problem will retain the mathematical form of $\,\bar{U}_{\it l}=\bar{k}_{
 \it l}\,\bar{W}_l\,$, with $\,\bar{k}_{\it l}\,$ keeping the same functional dependence on  $\,\rho\,$, $\,R\,$, and $\,\bar{\mu}\,$
 (or $\,\bar{J}\,$) as in (\ref{L4}), except that now $\,\mu\,$ and $\,{J}\,$ are equipped with overbars:
 \begin{subequations}
 \ba
 \bar{k}_{\it l}(\chi)
 &=&\frac{3}{2\,({\it l}\,-\,1)}\;\,\frac{\textstyle 1}{\textstyle 1\;+\;A_{\it l}\;\bar{\mu}(\chi)/\mu}
 \label{DD}\\
 \nonumber\\
 &=&\frac{3}{2\,({\it l}\,-\,1)}\;\,\frac{\textstyle 1}{\textstyle 1\;+\;A_{\it l}\;J/\bar{J}(\chi)}
 ~=~\frac{3}{2\,({\it l}\,-\,1)}\;\,\frac{\textstyle \bar{J}(\chi)}{\textstyle \bar{J}(\chi)\;+\;A_{\it l}\;J}~~~,~\quad~
  \label{k2bar}
 \ea
 \label{L335}
 \end{subequations}
 where
 \ba
 A_{\it l}\,\equiv~\frac{\textstyle{(2\,{\it{l}}^{\,2}\,+\,4\,{\it{l}}\,+\,3)\,\mu}}{\textstyle{{\it{l}}\,\mbox{g}\,\rho\,R}}\;=\;
 \frac{\textstyle{3\;(2\,{\it{l}}^{\,2}\,+\,4\,{\it{l}}\,+\,3)\,\mu}}{\textstyle{4\;{\it{l}}\,\pi\,G\,\rho^2\,R^2}}~=~\frac{\textstyle{3
 ~(2\,{\it{l}}^{\,2}\,+\,4\,{\it{l}}\,+\,3)~~J^{-1}}}{\textstyle{4\;{\it{l}}\,\pi\,G\,\rho^2\,R^2}}~~~,~~~
 \label{new}
 \ea
 Although expression (\ref{new}) for factors $\,A_l\,$ is very similar to expression (\ref{L4}) for their static counterparts, an
 important difference between (\ref{L4}) and (\ref{new}) should be pointed out. While in (\ref{L4}) we had the static (relaxed) rigidity
 and compliance, $\,\mu(\infty)\,$ and $\,J(\infty)=1/\mu(\infty)\,$, in (\ref{new}) the letters $\,\mu\,$ and $\,J\,$ may stand for any
 benchmark values satisfying $\,J=1/\mu\,$. This freedom stems from the fact that the products $\,A_l\,J\,$ entering (\ref{k2bar}) bear
 no dependence upon $\,J\,$ or $\,\mu\,$. The second term in the denominator of (\ref{DD}) contains $\,\bar{\mu}\,$. For convenience,
 we multiply and then divide $\,\bar{\mu}(\chi)\,$ by some $\,\mu\,$, and make the multiplier $\,\mu\,$ a part of $\,A_l\,$ as in
 (\ref{new}). This makes it easier for us to compare (\ref{new}) with its static predecessor (\ref{L4}). However the constant $\,\mu\,$
 in equations (\ref{L335}) and (\ref{new}) is essentially arbitrary, and is not obliged to coincide with, say, unrelaxed or relaxed rigidity. Accordingly, $\,J\,=\,1/\mu\,$
 is not obliged to be the unrelaxed or relaxed compliance.

 The above caveat is important, because in certain rheological models some of the unrelaxed or relaxed moduli may be zero or infinite.
 This will happen, for example, if we start with the Maxwell or Kelvin-Voigt body and perform a transition to a purely viscous medium.
 Fortunately, in realistic rheologies such things do not happen. Hence it will be convenient (and possible) to identify the $\,J\,$ from
 (\ref{new}) with the {\it{unrelaxed}} compliance $\,J=J(0)\,$ emerging in the rheological model (\ref{I64}). Accordingly, the rigidity
 $\,\mu=1/J\,$ from (\ref{new}) will be identified with the {\it{unrelaxed}} rigidity $\,\mu(0)=1/J(0)\,$. This convention will play a
 crucial role down the road, when we derive formula (\ref{kirha_1}).

 Writing the $~l\,$th$\,$ complex Love number as
 \ba
 \bar{k}_{\it{l}}(\chi)\;=\;{\cal{R}}{\it{e}}\left[\bar{k}_{\it{l}}(\chi)\right]\;+\;\inc\;
 {\cal{I}}{\it{m}}\left[\bar{k}_{\it{l}}(\chi)\right]\;=\;|\bar{k}_{\it{l}}(\chi)|\;
 e^{\textstyle{^{-\inc\epsilon_{\it l}(\chi)}}}
 \label{L36}
 \ea
 we express the phase lag $\,\epsilon_{\it l}(\chi)\,$ as:
 \ba
 |\bar{k}_{\it{l}}(\chi)|\;\sin\epsilon_{\it l}(\chi)\;=\;-\;{\cal{I}}{\it{m}}\left[\bar{k}_{\it{l}}(\chi)
 \right]\;\;\;.
 \label{ggffrr}
 \ea
 {\it{
 The importance of the products $\;|\bar{k}_{\it{l}}(\chi)|\;\sin\epsilon_{\it l}(\chi)\;$ lies in the fact that they show up in the
 terms of the Darwin-Kaula expansion of the tidal potential. As a result, it is these products, and not $\;k_{\it l}/Q\;$ as some
 think, which emerge in the expansions for tidal forces and torques, and for the dissipation rate.}}

 In an attempt to preserve the popular notation $\;k_{\it l}/Q\;$, one may {\it{define}} the inverse quality factor as the sine
 of the lag -- see the discussion in subsection \ref{damp}. In this case, though, one would have to employ the tidal lag $\,\epsilon_l\,
 $, and not the lag $\,\delta\,$ in the material (which we call the ``seismic" lag). Accordingly, one will have to write not
 $\;k_{\it l}/Q\;$ but $\;k_{\it l}/Q_l\;$, where $\,1/Q_l\,\equiv\,\sin\epsilon_l\,$.

 Importantly, the functional form of the frequency-dependence $\,\sin\epsilon_l(\chi)\,$ is different for different $\,l\,$. Thus an
 attempt of naming $\,\sin\epsilon_l\,$ as $\,1/Q\,$ would give birth to a whole array of different functions $\,Q_l(\chi)\,$. For a
 triaxial body, things will become even more complicated -- see footnote \ref{HH}. To conclude, it is not advisable to
 denote $\,\sin\epsilon_l\,$ with $\,1/Q\,$.

 It should be mentioned that the Darwin-Kaula theory of tides is equally applicable to tides in despinning and librating bodies. In all
 cases, the phase angle $\,\epsilon_l\,=\,\epsilon_l(\chi_{\textstyle{_{lmpq}}})\,$ parameterises the lag of the appropriate component
 of the bulge, while the absolute value of the complex Love number $\,|\bar{k}_l|\,=\,|\,\bar{k}_l(\chi_{\textstyle{_{lmpq}}})\,|\,$
 determines the magnitude of this component. The overall bulge being a superposition of these components, its height may vary in time.

 \subsection{The tangent of the tidal lag}

 In the denominator of (\ref{DD}) the term $\,1\,$ emerges due to self-gravitation, while $\,A_{\it l}\,J/\bar{J}(\chi)\,=
 \,A_{\it l}\,|\bar{\mu}(\chi)|/\mu\,$ describes how the bulk properties of the medium contribute to deformation and damping. So for a
 vanishing $\,A_{\it l}\,J/|\bar{J}(\chi)|\,$ we end up with the hydrostatic Love numbers $\,k_l=\,\frac{\textstyle 3}{\textstyle{2\,({
 \it l}\,-\,1)}}\,$, while the lag becomes nil, as will be seen shortly from (\ref{L39b}). On the contrary, for very large $\,A_{\it l}
 \,J/\bar{J}(\chi)\,$, we expect to obtain the Love numbers and lags ignorant of the shape of the body.

 To see how this works out, combine formulae (\ref{112}) and (\ref{k2bar}), to arrive at
 \begin{subequations}
 \ba
 \tan \epsilon_{\textstyle{_l}}\,=~-~\frac{{\cal{I}}{\it{m}}\left[\bar{k}_{\it{l}}(\chi)\right]}{{\cal{R}}{\it{e}}\left[\bar{k}_{\it{l}}
 (\chi)\right]}~=~-~\frac{\,A_l~J~\,{\cal{I}}{\it{m}}\left[\bar{J}(\chi)\right]}{\,\left\{\,{\cal{R}}{\it{e}}\left[\bar{J}(\chi)\right]
 \,\right\}^2\,+~\left\{\,{\cal{I}}{\it{m}}\left[\bar{J}(\chi)\right]\,\right\}^2
 \,+~A_l~J~{\cal{R}}{\it{e}}\left[\bar{J}(\chi)\right]\,}~=\quad\quad\quad\quad
  \label{kirha_1a}
 \ea
 \ba
 \frac{A_l~\left[\,\zeta~z^{-1}\,+~z^{-\alpha}~\sin\left(\,\frac{\textstyle\alpha\,\pi}{\textstyle 2}\,\right)~\Gamma(1+\alpha)\,\right]}{
 \left[1+z^{-\alpha}\cos\left(\frac{\textstyle\alpha\pi}{\textstyle 2}\right)\Gamma(1+\alpha)\right]^2+
 \left[\zeta z^{-1}+z^{-\alpha}\sin\left(\frac{\textstyle\alpha\pi}{\textstyle 2}\right)\Gamma(1+\alpha)\right]^2+A_l
 \left[1+z^{-\alpha}\cos\left(\frac{\textstyle\alpha\pi}{\textstyle 2}\right)\Gamma(1+\alpha)\right]}\quad,\quad
  \label{kirha_1b}
 \ea
 \label{kirha_1}
 \end{subequations}
 $\,z\,$ being the dimensionless frequency defined by (\ref{sim}).

 Comparing this expression with expression (\ref{LL45}), over different frequency bands, we shall be able to explore how the tidal lag
 $\,\epsilon_{\textstyle{_l}}\,$ relates to the lag in the material $\,\delta\,$ (the ``seismic lag").

 While expression (\ref{kirha_1b}) is written for the Andrade model, the preceding formula (\ref{kirha_1a}) is general and works for an
 arbitrary linear rheology.

 \subsection{The negative imaginary part of the complex Love number}\label{last}

 As we already mentioned above, rheology influences the tidal behaviour of a planet through the following sequence of steps. A
 rheological model postulates the form of $\,\bar{J}(\chi)\,$. This function, in its turn, determines the form of $\,\bar{k}_{\it{l}}
 (\chi)\,$, while the latter defines the frequency dependence of the products $\,|\bar{k}_{\it{l}}(\chi)|\;\sin\epsilon(\chi)\,$ which
 enter the tidal expansions.

 To implement this concatenation, one has to express $\,|\bar{k}_{\it{l}}(\chi)|\;\sin\epsilon(\chi)\,$ via $\,\bar{J}(\chi)\,$. This
 can be done by combining (\ref{L335}) with (\ref{ggffrr}). It renders:
 \ba
 |\bar{k}_{\it l}(\chi)|\;\sin\epsilon_{\it l}(\chi)\;=\;-\;{\cal{I}}{\it{m}}\left[\bar{k}_{\it l}(\chi)
 \right]\;=\;\frac{3}{2\,({\it l}\,-\,1)}\;\,\frac{-\;A_l\;J\;{\cal{I}}{\it{m}}\left[\bar{J}(\chi)\right]}{\left(\;{\cal{R}}{
 \it{e}}\left[\bar{J}(\chi)\right]\;+\;A_l\;J\;\right)^2\;+\;\left(\;{\cal{I}}{\it{m}}
 \left[\bar{J}(\chi)\right]\;\right)^2} ~~~,~~~~~
 \label{L39b}
 \ea
 a quantity often mis-denoted\footnote{~One can write the left-hand side of (\ref{L39b}) with $\,k_l/Q\,$ only if the quality factor is defined through
 (\ref{pref}) and endowed with the subscript $\,l\,$.} as $\,k_l/Q\,$. Together, formulae (\ref{112}) and (\ref{L39b}), give us the
 frequency dependencies for the factors $\,|\bar{k}_{\it l}(\chi)|~\sin\epsilon_{\it l}(\chi)\,$ entering the theory of bodily tides.
 For detailed derivation of those dependencies, see Efroimsky (2012).

 As explained in Section \ref{ka}, employment of expressions (\ref{ggffrr} - \ref{L39b}) needs some care. Since both $\,\bar{U}\,$ and $\,
 \bar{k}_{l}\,$ are in fact functions not of the forcing frequency $\,\chi\,$ but of the tidal mode $\,\omega\,$, formulae (\ref{ggffrr} -
 \ref{L39b})  should be equipped with multipliers ~sgn$\,\omega_{\textstyle{_{lmpq}}}\,$, when plugged into the expression for the $lmpq$
 component of the tidal torque. With this important caveat in mind, and with the subscripts $\,lmpq\,$ reinstalled, the complete
 expression will read:
  \ba
 |\bar{k}_{\it l}(\chi_{\textstyle{_{lmpq}}})|\;\sin\epsilon_{\it l}(\chi_{\textstyle{_{lmpq}}})\,=\,
 \frac{3}{2\,({\it l}-1)}\;\,\frac{-\;A_l\;J\;{\cal{I}}{\it{m}}\left[\bar{J}(\chi_{\textstyle{_{lmpq}}})\right]}{\left(\,{\cal{R}}{
 \it{e}}\left[\bar{J}(\chi_{\textstyle{_{lmpq}}})\right]\,+\,A_l\,J\,\right)^2+\,\left(\,{\cal{I}}{\it{m}}
 \left[\bar{J}(\chi_{\textstyle{_{lmpq}}})\right]\,\right)^2}~~\mbox{sgn}\,\omega_{\textstyle{_{lmpq}}}~~,~~~~
 \label{L39bbb}
 \ea
 a general formula valid for an arbitrary linear rheological model.

 To make use of this and other formulae, it would be instructive to estimate the values of $\,A_l\,$ for terrestrial objects
 of different size. In Table \ref{table}, we present estimates of $\,A_2\,$ for Iapetus, Mars, solid Earth, a hypothetical solid
 superearth having a density and rigidity of the solid Earth and a radius equal to 2 terrestrial radii ($\,R=2R_{\textstyle{_{\bigoplus}}}\,$), and also a twice larger
 hypothetical superearth ($\,R=4R_{\textstyle{_{\bigoplus}}}\,$) of the same rheology.

 \begin{table*}[htdp]\caption{~Estimates of $\,A^{\textstyle{^{(static)}}}_2\,$ for rigid celestial bodies. The values of
  $\,A^{\textstyle{^{(static)}}}_2\,$ are calculated using equation (\ref{A2}) and are rounded to the second figure.
 %
 }
 \begin{center}
 \begin{tabular}{lcccc}
 \hline
 \hline
                 & radius $\,R\,$      & ~mean density $\rho\,$         & mean relaxed  & the $\,$resulting\\
                 &                     &                                & ~shear rigidity $\mu(\infty)$   & ~estimate for $A_2\,$\\
 \hline
 \hline
                 &                     &                                &                         &                        ~\\
 Iapetus         & ~$7.4\times10^5$ m  & ~~$1.1\times10^3$ kg/m$^{\,3}$ & $4.0\times10^9\,$ Pa    &  $200$                   ~\\
                 &                     &                                &                         &                        ~\\
 \hline\\
 Mars            & ~$3.4\times10^6$ m  & ~~$3.9\times10^3$ kg/m$^{\,3}$ & $1.0\times10^{11}\,$ Pa &   $19$                   ~\\
 ~\\
 \hline\\
 The Earth       & ~$6.4\times10^6$ m & ~~$5.5\times10^3$ kg/m$^{\,3}$ & $0.8\times10^{11}\,$ Pa & $2.2$                    ~\\
 ~\\
 \hline
 A $\,$hypothetical $\,$superearth
                 &                    &                                 &                            &                        ~\\
 with $~~R\,=\,2\,R_{\textstyle{_{\bigoplus}}}~~$ and the
                 & ~$4.5\times10^8$ m & ~~$5.5\times 10^3$ kg/m$^{\,3}$ & $0.8\times 10^{11}\,$ Pa & $0.55$
 ~\\
 same rheology as the Earth
                 &                    &                                 &                            &                        ~\\
  ~\\
 \hline
 A $\,$hypothetical $\,$superearth
                 &                    &                                 &                            &                        ~\\
 with $~~R\,=\,4\,R_{\textstyle{_{\bigoplus}}}~~$ and the
                 & ~$9.0\times10^8$ m & ~~$5.5\times 10^3$ kg/m$^{\,3}$ & $0.8\times 10^{11}\,$ Pa & $0.14$
 ~\\
 same rheology as the Earth
                 &                    &                                 &                            &                        ~\\

 \hline
 \hline
 \end{tabular}
 \end{center}
 \label{table}
 \end{table*}
 Taken the uncertainty of structure and the roughness of our estimate, all quantities in the table have been rounded to the first decimal.
 The values of Iapetus' and Mars' rigidity were borrowed from Castillo-Rogez et al. (2011) and Johnson et al. (2000), correspondingly.

\begin{figure}[]
 \begin{center}
 \vspace{1cm}
  \includegraphics[width=17.9cm
  ]{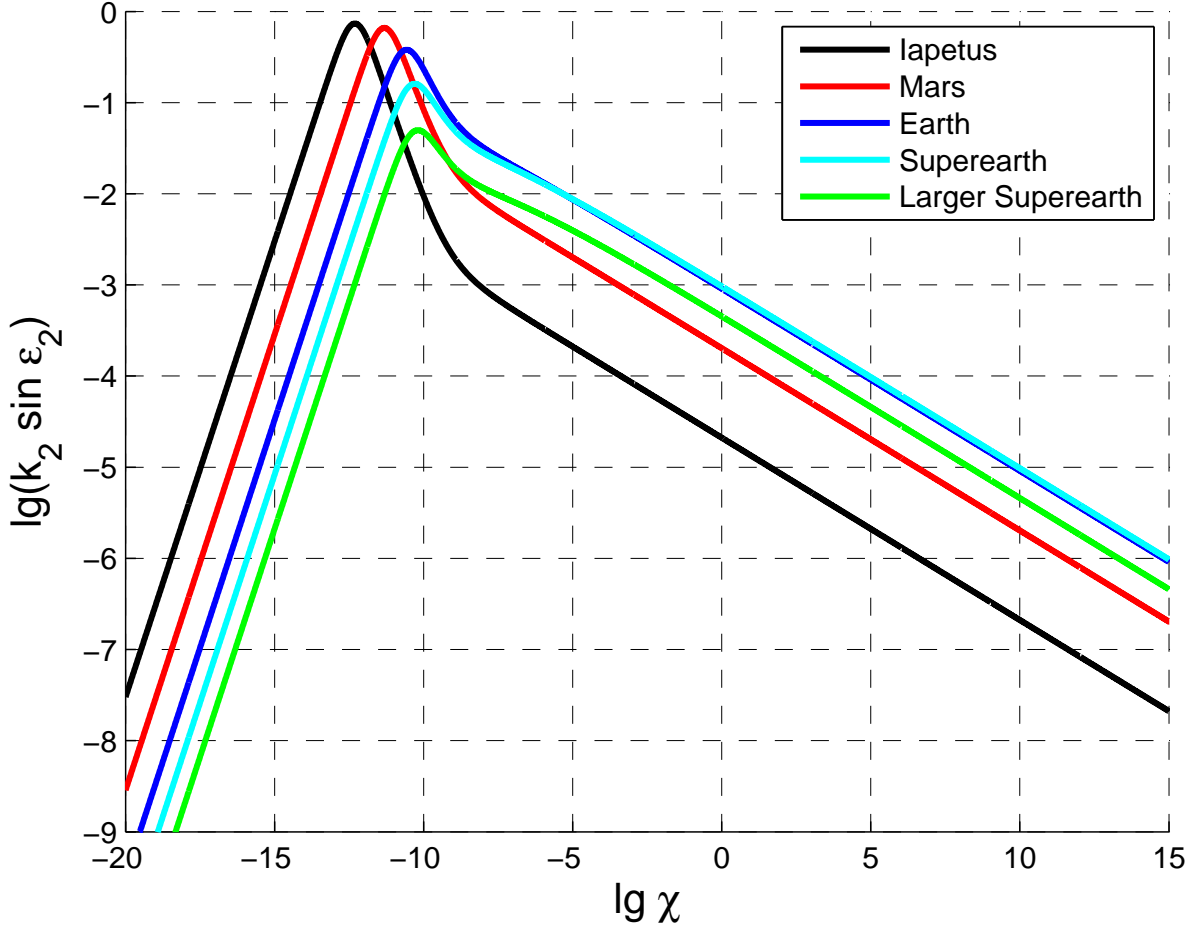}
 \end{center}
 \caption{\small{
 Negative imaginary part of the complex quadrupole Love number, $\,k_2\,\sin\epsilon_2=\,-\,{\cal{I}}{\it{m}}\left[\bar{k}_2(\chi)
 \right]\,$, as a function of the tidal frequency $\chi\,$. The black, red, and blue curves refer, correspondingly, to Iapetus, Mars,
 and the solid Earth. The cyan and green curves refer to the two hypothetical superearths described in Table~\ref{table}. These superearths
 have the same rheology as the solid Earth, but have sizes $~R\,=\,2\,R_{\textstyle{_{\bigoplus}}}~$ and $~R\,=\,4\,
 R_{\textstyle{_{\bigoplus}}}~$. Each of these five objects is modeled with a homogeneous near-spherical self-gravitating Andrade body
 with $\,\alpha=0.2\,$ and $\,\tau_{_A}=\tau_{_M}=10^{10}\,$s. In the limit of vanishing tidal frequency $\chi$, the factors $\,k_2\,
 \sin\epsilon_2\,$ approach zero, which is natural from the physical point of view. Indeed, an $\,lmpq\,$ term in the expansion for
 tidal torque contains the factor $~k_l(\chi_{\textstyle_{lmpq}})\,\sin\epsilon_l(\chi_{\textstyle_{lmpq}})\,$. On crossing the
 $\,lmpq\,$ resonance, where the frequency $\chi_{\textstyle_{lmpq}}$ goes through zero, the factor $\,k_l(\chi_{
 \textstyle_{lmpq}})\,\sin\epsilon_l(\chi_{\textstyle_{lmpq}})\,$ must vanish, so that the $\,lmpq\,$ term of the
 torque could change its sign.}}
 \label{Figure_2}
 \end{figure}

 In Figure \ref{Figure_2}, we compare the behaviour of $\,k_2\,\sin\epsilon_2=|\bar{k}_{2}(\chi)|\,\sin\epsilon_{2}(\chi)\,$
 for the values of $\,A_2\,$ appropriate to Iapetus, Mars, solid Earth, and hypothetical superearths with
 $~R\,=\,2\,R_{\textstyle{_{\bigoplus}}}~$ and $~R\,=\,4\,R_{\textstyle{_{\bigoplus}}}~$ , as given in Table \ref{table}.
   Self-gravitation pulls the tides down, mitigating their magnitude and the value of the tidal torque. Hence, the heavier the body the
 lower the appropriate curve. This rule is observed well at low frequencies (the viscosity-dominated range). In the intermediate zone
 and in the high-frequency band (where inelastic creep dominates friction), this rule starts working only for bodies larger than about
 twice the Earth size. If we fix the tidal frequency at a sufficiently high value, we shall see that the increase of the size from
 that of Iapetus to that of Mars and further to that of the Earth results in an {\it{increase}} of the intensity of the tidal
 interaction. For a $~R\,=\,2\,R_{\textstyle{_{\bigoplus}}}~$ superearth, the tidal factor $\,k_2\,\sin\epsilon_2\,$ is about the
 same as that for the solid Earth, and begins to decrease for larger radii (so the green curve for the larger superearth is located
 fully below the cyan curve for a smaller superearth).

 \section{Tidal dissipation versus seismic dissipation,\\
 in the inelasticity-dominated band}

 In this section, we shall address only the higher-frequency band of the spectrum, i.e., the range where inelasticity dominates
 viscosity and the Andrade model is applicable safely. Mind though that the Andrade model can embrace also the near-Maxwell
 behaviour, and thus can be applied to the low frequencies, provided we ``tune" the dimensionless parameter $\,\zeta\,$ appropriately
 -- see subsection {\ref{ggg}} above.

 \subsection{Response of a sample of material}

 At frequencies higher than some threshold value $\,\chi_0\,$, dissipation in minerals is mainly due to inelasticity rather than to
 viscosity.\footnote{~For the solid Earth, this threshold is about 1 yr$^{-1}$ (Karato \& Spetzler 1990). Being temperature sensitive, the
 threshold may assume different values for other terrestrial planets.

 Also mind that the transition is not sharp and can extend over a decade or more.} Hence at these frequencies $\,\zeta\,$ should be of
 order unity or smaller, as can be seen from (\ref{compliance_2}). This entails two consequences. First, the condition $\,\chi\gg{
 \textstyle 1}/{(\textstyle\zeta\,\tau_{_M})}~$, ~i.e., $\,z\gg1\,$ is obeyed reliably, for which reason the first term dominates the denominator in (\ref{LL45}). Second,
 either the condition $\,z\,\gg\,1\,$ is stronger than $\,z\,\gg\,\zeta^{\textstyle{^{\textstyle{\frac{1}{1-\alpha}}}}}\,$ or the two
 conditions are about equivalent. Hence the inelastic term dominates the numerator in (\ref{LL45}): $\,~z^{-\alpha}\,\gg\,z^{-1}\,\zeta~$.

 Altogether, over the said frequency range, (\ref{LL45}) simplifies to:
 \ba
 \tan \delta~\approx~(\chi\,\tau_{_A})^{\textstyle{^{-\alpha}}}\sin\left(\frac{\alpha\,\pi}{2}\right)~\Gamma(\alpha+1)~=~
 \left(\chi\,\zeta\,\tau_{_M}\right)^{\textstyle{^{-\alpha}}}\sin\left(\frac{\alpha\,\pi}{2}\right)~\Gamma(\alpha+1)
 ~~~.~~\,~
 \label{LL4433a}
 \ea
 Clearly $\,\tan\delta\ll 1\,$, wherefore $\,\tan\delta\approx\sin\delta\approx\delta\,$. For the seismic quality factor, we then have:
 \ba
 {}^{(seismic)}Q^{-1}~\approx~(\chi\,\tau_{_A})^{\textstyle{^{-\alpha}}}\sin\left(\frac{\alpha\,\pi}{2}\right)~
 \Gamma(\alpha+1)~~~,
 \label{andrade_1}
 \label{LL43a}
 \ea
 no matter which of the three definitions (\ref{DI6} - \ref{Goldreich}) we accept. Be mindful, that here we use the term {\it{seismic}}
 broadly, applying it also to a sample in a lab.

 \subsection{Tidal response of a homogeneous near-spherical body}

 Recall that defect unpinning stays effective at frequencies above some threshold $\,\chi_0\,$, which is likely to be above or, at
 least, not much lower than the inverse Maxwell time.\footnote{~Dislocations may break away from the pinning agents (impurities,
 nodes, or jogs), or the pinning agents themselves may move along with dislocations. These two processes are called ``unpinning",
 and they go easier at low frequencies, as the energy barriers become lower (Karato \& Spetzler 1990, ~section 5.2.3).} E.g., for
 the solid Earth, $\,\chi_{0}\sim 1\,$yr$^{-1}\,$ while $\,\tau_{_M}\,\sim \,500\,$yr. Over this frequency band, the free parameter
 $\,\zeta\,$ may be of order unity or slightly less than that. (This parameter grows as the frequencies become short of $\,\chi_0\,$.)
 Under these circumstances, in equation (\ref{kirha_1}) we have: $~\zeta z^{-1}\ll z^{-\alpha} \ll 1~$, whence equation (\ref{kirha_1})
 becomes:
 \ba
 \tan\epsilon_{\textstyle{_l}}\,\approx~\frac{A_l}{1\,+\,A_l}~z^{-\alpha}~\sin\left(\frac{\textstyle\alpha\pi}{\textstyle 2}\right)~
 \Gamma(1+\alpha)~~~.
 \label{Earth_1}
 \ea
 In combination with (\ref{LL4433a}), this renders:
 \ba
 \tan\epsilon_{\textstyle{_l}}~=~\frac{A_l}{1\,+\,A_l}~\tan\delta~~~.
 \label{Earth}
 \ea
 Had we defined the quality factors as cotangents, like in (\ref{Goldreich}), then we would have to conclude from (\ref{Earth}) that the
 tidal and seismic quality factors coincide for small objects (with $\,A_l\gg 1\,$) and differ very considerably for superearths (i.e.,
 for $\,A_l\ll 1\,$). Specifically, the so-defined quality factor $\,Q_{\textstyle{_l}}\,$ of a superearth would be larger than its
 seismic counterpart $\,Q\,$ by a factor of about $\,A_l^{-1}\,$.

 In reality, the quality factors should be used for illustrative purposes only, because practical calculations involve the factor
 $\,|\bar{k}_{\textstyle{_l}}(\chi_{\textstyle{_{lmpq}}})|\,\sin\epsilon_{\textstyle{_l}}(\chi_{\textstyle{_{lmpq}}})\,$ rendered by
 (\ref{L39bbb}). It is this factor which enters the $\,lmpq\,$ term of the Fourier expansion of tides. Insertion of (\ref{A3} -
 \ref{A4}) into (\ref{L39bbb}) furnishes the following expression valid in the inelasticity-dominated band:
 \ba
 |\bar{k}_{\textstyle{_l}}(\chi_{\textstyle{_{lmpq}}})|\,\sin\epsilon_{\textstyle{_l}}(\chi_{\textstyle{_{lmpq}}})\,\approx
 \quad\quad\quad\quad\quad\quad\quad\quad\quad\quad\quad\quad\quad\quad\quad\quad\quad\quad\quad\quad\quad\quad\quad\quad
 \quad\quad\quad\quad\quad\quad\quad\quad\quad\quad\quad\quad\quad\quad\quad\quad
 \nonumber
 \ea
 \ba
 \quad\quad\frac{3}{2\,(l-1)}\;\frac{A_{\textstyle{_l}}}{(A_{\textstyle{_l}}+\,1)^2}~\sin\left(\frac{\alpha\pi}{2}\right)~\Gamma(\alpha+1)
 ~\,\zeta^{-\alpha}\,\left(\tau_{_M}~\chi_{\textstyle{_{lmpq}}}\right)^{-\alpha}\,\mbox{sgn}\,\omega_{\textstyle{_{lmpq}}}\,\quad,
 \quad\mbox{for}\quad\chi_{\textstyle{_{lmpq}}}\,\gg\,\chi_{_{HI}}~~,\quad
 \label{aaaa}
 \ea
 $\chi_{_{HI}}\,$ being the boundary between the high and intermediate frequencies, i.e., between the inelasticity-dominated band and the
 transitional zone. Expression (\ref{aaaa}) resembles the frequency dependency for $\,|\bar{J}(\chi)|\,\sin\delta(\chi)\,=\,-\,{\cal I}
 {\it m}[\bar{J}(\chi)]~$ at high frequencies (see equation \ref{A3}). In Figure 2, dependency (\ref{aaaa}) corresponds to the slowly
 descending slope on the far right.

 A detailed derivation of (\ref{aaaa}) from formulae (\ref{A3} - \ref{A4}) and (\ref{L39bbb}) is presented in the Appendix to Efroimsky
 (2012). For terrestrial objects several times smaller than the Earth (so $\,A_{\textstyle{_{l}}}\gg 1\,$), the threshold turns out to be
 \ba
 \chi_{{_{HI}}}\,=\,\tau_{_M}^{-1}\,\zeta^{\textstyle{^{\textstyle\,\frac{\alpha}{1-\alpha}}}}~~~.
 \label{es1}
 \ea
 For superearths (i.e., for $\,A_{\textstyle{_{l}}}\ll 1\,$), the threshold becomes
 \ba
 \chi_{{_{HI}}}\,=\,\tau_{_A}^{-1}\,=\,\tau_{_M}^{-1}\,\zeta^{-1}~~~.
 \label{es2}
 \ea
 Near the borderline between the inelasticity-dominated band and the transitional zone, the parameter $\,\zeta\,$ could be of order
 unity. It may as well be lower than unity, though not much (hardly by an order of magnitude), because too low a value of $\,\zeta\,$
 would exclude viscosity from the play completely. We however expect viscosity to be noticeable near the transitional zone.

 Finally, it should be reiterated that at frequencies lower than some $\,\chi_{0}\,$ the defect-unpinning process becomes less effective,
 so inelasticity becomes less effective than viscosity, and the free parameter $\,\zeta\,$ begins to grow. Hence, if the above estimates
 for $\,\chi_{{_{HI}}}\,$ turn out to be lower than $\,\chi_{0}\,$, we should set $\,\chi_{{_{HI}}}\,=\,\chi_{0}\,$ ``by hand".

 \section{Tidal dissipation versus seismic dissipation,\\
 in the viscosity-dominated band}

 When frequency $\,\chi\,$ becomes short of some $\,\chi_{\textstyle{_0}}\,$, the rate of defect-unpinning-caused inelastic dissipation
 decreases and viscosity begins to take over inelasticity.

 If we simply assume the free parameter $\,\zeta\,$ to be of order unity everywhere, i.e., assume that the Maxwell and Andrade timescales are
 everywhere comparable, then application of the Andrade model will set $\,\chi_{\textstyle{_0}}\,$ to be of order $\,\tau^{-1}_{_M}\,$.
 Anelasticity will dominate at frequencies above that threshold, while below it the role of viscosity will be higher. This approach
 however would be simplistic, because the actual location of the threshold should be derived from microphysics and may turn out to
 differ from $\,\tau^{-1}_{_M}\,$ noticeably. For example, in the terrestrial mantle the transition takes place at frequencies as high
 as 1 $yr^{-1}$ (Karato \& Spetzler 1990) and may be spread over a decade or more into lower frequencies, as we shall see from equation
 (\ref{condition}).

 Another somewhat simplistic option would be to assume that $\,\zeta\sim 1\,$ at frequencies above $\,\chi_{\textstyle{_0}}\,$, and to
 set $\,\zeta=\infty\,$ at the frequencies below $\,\chi_{\textstyle{_0}}\,$. The latter would be equivalent to claiming that below this
 threshold the mantle is described by the Maxwell model. In reality, here we are just entering a transition zone, where $\,\zeta\,$
 increases with the decrease of the frequency. While it is clear that in the denominator of (\ref{LL45}) the first term dominates, the
 situation with the numerator is less certain. Only after the condition
 \ba
 \zeta\,\gg\,(\chi\,\tau_{_M})^{\textstyle{^{\textstyle\frac{1-\alpha}{\alpha}}}}\,\approx~
 (\chi\,\tau_{_M})^4
 \label{condition}
 \ea
 is obeyed, the viscous term $\,1/(\chi\,\tau_{_M})\,$ becomes leading. This way, although
 $\,\zeta\,$ begins to grow as the frequency decreases below $\,\chi_0\,$, the frequency may
 need to decrease by another decade or more before threshold (\ref{condition}) is reached.

 \subsection{Response of a sample of material}

 Accepting the approximation that the transition zone is narrow$\,$\footnote{~For a broader transition zone, the rheology will approach
 that of Maxwell at lower frequencies. This though will not influence our main conclusions.} and that the predominantly viscous regime
 is reached already at $\,\chi_0\,$ or shortly below, we approximate the tangent of the lag with
 \ba
 \tan\delta~\approx~(\chi\,\tau_{_M})^{-1}~~~,
 \label{dell}
 \ea
 whence
 \ba
  \sin\delta~\approx~
 \left\{
 \begin{array}{c}
 \quad   (\chi\,\tau_{_M})^{-1}\quad~\mbox{for}\quad   \tau_{_M}^{-1}\,\ll\,\chi\,\ll\,\chi_{\textstyle{_0}}~~~,\\
 ~\\
 \quad\quad 1\quad\quad\,\quad\mbox{for}\quad   0\,\leq\,\chi\,\ll\,\tau_{_M}^{-1}~~~.
 \end{array}{}
 \right.
 \label{vcv}
 \ea

 \subsection{Tidal response of a homogeneous near-spherical body}

 When viscosity dominates inelasticity, expression (\ref{kirha_1}) gets reduced to the following form:
 \ba
 \tan\epsilon_{\textstyle{_l}}\,\approx~\frac{A_l}{1\,+\,A_l\,+\,(\zeta\,z^{-1})^{\textstyle{^2\,}}}~\,\zeta\,z^{-1}~=~
 ~\frac{A_l}{1\,+\,A_l\,+\,\left(\textstyle{\chi\,\tau_{_M}}\right)^{{{-\,2}}}\,}~\,\frac{1}{\chi\,\tau_{_M}}~~~,
 \label{gigg}
 \ea
 comparison whereof with (\ref{dell}) renders:
 \ba
 \tan\epsilon_{\textstyle{_l}}\,\approx~\frac{A_l}{1\,+\,A_l\,+\,\left(\textstyle{\chi\,\tau_{_M}}\right)^{-2}\,}~\,\tan\delta~=
 ~\frac{A_l}{1\,+\,A_l\,+\,\tan^2\delta\,}~\,\tan\delta~~~.
 \label{Earth'}
 \ea
 Now two special cases should be considered separately.

 \subsubsection{Small bodies and small terrestrial planets}

 As illustrated by Table \ref{table}, small bodies and small terrestrial planets have $\,A_l\gg 1\,$. So formulae
 (\ref{gigg}) and (\ref{Earth'}) take the form of
 \ba
 \tan\epsilon_{\textstyle{_l}}\,~\approx~
 \left\{
 \begin{array}{c}
 \quad\quad \frac{\textstyle 1}{\textstyle
 \chi\,\tau_{_M}}\quad~\quad~\quad\quad\mbox{for}\quad\quad\frac{\textstyle 1}{\textstyle{\tau_{_M}}\,\sqrt{A_l+1}}\,\ll\,\chi\ll\chi_{\textstyle{_0}}~
 ~~,\quad\quad\quad\quad\quad\quad\quad\quad
 \label{90a}\\
 ~\\
 \quad\quad  ~\textstyle A_l\,\chi\,\tau_{_M}~\quad\quad\quad\quad~\mbox{for}\quad\quad 0\,\leq\,\chi\ll\,\frac{\textstyle
 1}{\textstyle{\tau_{_M}}\,\sqrt{A_l+1}}~\approx~\frac{\textstyle 1}{\textstyle{\tau_{_M}}\,\sqrt{A_l}}~~~,
 \quad~\quad\,\quad\quad
 \label{90b}
 \end{array}
 \right.
 \label{90}
 \ea
  and
 \ba
 \tan\epsilon_{\textstyle{_l}}\,~\approx~
 \left\{
 \begin{array}{c}
 \frac{\textstyle A_l}{\textstyle{1\,+\,A_l}}~\tan\delta~\approx~\tan\delta\quad\quad\quad\mbox{for}\quad\quad\frac{\textstyle 1}{\textstyle{
 \tau_{_M}}\,\sqrt{A_l+1}}\ll \chi\ll\chi_{\textstyle{_0}}~~~,\quad\,\quad\quad\quad\quad\quad\\
 ~\\
 \quad\quad \quad\quad\quad\quad ~\frac{\textstyle A_l}{\,\textstyle \tan\delta\,}~\quad\quad\quad\quad\quad\mbox{for}\quad\quad 0\,\leq\,\chi\ll
 \frac{\textstyle 1}{\textstyle{\tau_{_M}}\,\sqrt{A_l+1}}~\approx~\frac{\textstyle 1}{\textstyle{\tau_{_M}}\,\sqrt{A_l}}~~~.\quad~\quad
 \end{array}
 \right.
 \label{91}
 \ea

 Had we defined the quality factors as cotangents of $\,\epsilon_{\textstyle{_l}}\,$ and $\,\delta\,$, we would be faced with a situation
 that may at first glance appear embarrassing: in the zero-frequency limit, the so-defined tidal $\,Q_{\textstyle{_l}}\,$ would become
 {\it{inversely}} proportional to the so-defined seismic $\,Q\,$ factor. This would however correspond well to an obvious physical fact:
 when the satellite crosses the $\,lmpq\,$ commensurability, the $\,lmpq\,$ term of the average tidal torque acting on a satellite must
 smoothly pass through nil, together with the $\,lmpq\,$ tidal mode. (For example, the orbital average of the principal tidal torque
 $\,lmpq\,=\,2200\,$ must vanish when the satellite crosses the synchronous orbit.) For a more accurate explanation in terms of the
 $\,|\bar{k}_l(\chi)|\,\sin\epsilon_l(\chi)\,$ factors see subsection \ref{be} below.

 \subsubsection{Superearths}

 For superearths, we have $\,A_l\ll 1\,$, so (\ref{Earth'}) becomes
 \ba
 \tan\epsilon_{\textstyle{_l}}\,~\approx~A_l\,\chi\,\tau_{_M}~=~\frac{\textstyle A_l}{\,\textstyle \tan\delta\,}~\quad\quad
 \quad\quad\mbox{for}\quad\quad 0\leq\chi\ll\frac{\textstyle 1}{\textstyle{\tau_{_M}}\,\sqrt{A_{\textstyle{_l}}+1}}~\approx
 ~\frac{\textstyle 1}{\textstyle{\tau_{_M}}} ~~.\,~~~
 \label{sup}
 \ea
 Here we encounter the same apparent paradox: had we defined the quality factors as cotangents of $\,
 \epsilon_{\textstyle{_l}}\,$ and $\,\delta\,$, we would end up with a tidal
 $\,Q_{\textstyle{_l}}\,$ {{inversely}} proportional to its seismic counterpart $\,Q\,$.
 %
 A qualitative explanation to this ``paradox" is the same as in the subsection above, a more accurate elucidation to be given shortly
 in subsection \ref{be} .

 Another seemingly strange feature is that in this case (i.e., for $\,A_l\ll 1\,$) the tangent of the tidal lag skips the range of
 inverse-frequency behaviour and becomes linear in the frequency right below the inverse Maxwell time. This however should not surprise
 us, because the physically meaningful products $\,k_l\,\sin\epsilon_l\,$
 still retain a short range over which they demonstrate the inverse-frequency behaviour. This can be understood from Figure
 \ref{Figure_2}. There, on each plot, a short segment to the right of the maximum corresponds to the situation when
 $\,k_l\,\sin\epsilon_l\,$ scales as inverse frequency -- see formula (\ref{bbbb}) below.

 Thus we once again see that the illustrative capacity of the quality factor is limited.
 To spare ourselves of surprises and ``paradoxes", we should always keep in mind that the
 actual calculations are based on the frequency dependence of $\,|\bar{k}_{\textstyle{_l}}(
 \chi_{\textstyle{_{lmpq}}})|\,\sin\epsilon_{\textstyle{_l}}(\chi_{\textstyle{_{lmpq}}})\,$.

 \subsection{Tidal response in terms of $\,|\bar{k}_{\textstyle{_l}}(\chi)|\,\sin\epsilon_{\textstyle{_l}}(\chi)\,$}\label{be}

 Combining (\ref{A3} - \ref{A4}) with (\ref{L39bbb}), one can demonstrate that in the intermediate-frequency zone the tidal factors
 scale as
 \ba
 |\bar{k}_{\textstyle{_l}}(\chi)|\,\sin\epsilon_{\textstyle{_l}}(\chi)\,\approx\,\frac{3}{2\,(l-1)}\;\frac{A_{\textstyle{_l}}}{(A_{
 \textstyle{_l}}+1)^2}\,~
 \left(\,\tau_{_M}\,\chi\,\right)^{-1}\quad,\quad~\quad\mbox{for}\quad\quad\tau_{_M}^{-1}\gg\chi\gg
 \tau_{_M}^{-1}\,(A_{\textstyle{_l}}+1)^{-1}~\,~,\quad\quad
  \label{bbbb}
 \ea
 which corresponds to the short segment on the right of the maximum on Figure 2.

 From the same formulae (\ref{A3} - \ref{A4}) and (\ref{L39bbb}), it ensues that the low-frequency behaviour looks as
  \ba
 |\bar{k}_{\textstyle{_l}}(\chi)|~\sin\epsilon_{\textstyle{_l}}(\chi)\,\approx\,\frac{3}{2\,(l-1)\textbf{}}~{A_{\textstyle{_l}}}~\,
 \tau_{_M}~\chi\quad\quad,\quad\quad\quad\quad~\quad\quad\,\mbox{for}~\quad~\quad\,\tau_{_M}^{-1}\,(A_{\textstyle{_l}}+1)^{
 -1}\,\gg\,\chi~~~,\quad\quad~\quad
 \label{cccc}
 \ea
 a regime illustrated by the slope located on the left of the maximum on Figure 2.

 Details of derivation of (\ref{bbbb}) and (\ref{cccc}) can be found in the Appendix to Efroimsky (2012).

 Just as expression (\ref{aaaa}) resembled the frequency dependency (\ref{A3}) for $\,|\bar{J}
 (\chi)|\,\sin\delta(\chi)
 ~$ at high frequencies, so (\ref{bbbb}) resembles the behaviour of $\,|\bar{J}(\chi)|\,\sin
 \delta(\chi)
 ~$ at low frequencies. At the same time, (\ref{cccc}) demonstrates a feature inherent only in tides,
 and not in the behaviour of a sample of material: at $\,\chi<\tau_{_M}^{-1}(A_{\textstyle{_l}}+1)^{-1}\,=\,\frac{\textstyle\mu}{
 \textstyle\eta}\,(A_{\textstyle{_l}}+1)^{-1}$, the factor $\;|\bar{k}_{\it{l}}(\chi)|\;\sin\epsilon_{\it{l}}(\chi)\;$ becomes linear
 in $\,\chi\,$. This is not surprising, as the $\,lmpq\,$ component of the average tidal torque or force must pass smoothly through zero and
 change its sign when the $\,lmpq\,$ commensurability is crossed (and the $lmpq$ tidal mode goes through zero and changes sign).

  \section{Why the $\,lmpq\,$ component of the tidal torque does not scale as $\,R^{{{\,2l+1}}}$}

 A Fourier component $\,{\cal{T}}_{\textstyle{_{lmpq}}}\,$ of the tidal torque acting on a perturbed primary is proportional to
 $\,R^{\textstyle{^{\,2l+1}}}\,k_l\,\sin\epsilon_l\,$, where $\,R\,$ is the primary's mean equatorial radius. Neglect of the $\,R-$dependence of
 the tidal factors $\,k_l\,\sin\epsilon_l\,$ has long been source of misunderstanding on how the torque scales with the radius.

 From formulae (\ref{aaaa}) and (\ref{bbbb}), we see that everywhere except in the closest vicinity of the resonance the tidal factors
 are proportional to $~A_l/(1+A_l)^2~$ where $\,A_l\sim R^{-2}\,$ according to (\ref{new}). Thence
 the overall dependence of the tidal torque upon the radius becomes:
 \ba
  \nonumber
  \mbox{Over the frequency band}\quad\chi\,\gg\,\tau^{-1}_{_M}\,\left(1\,+\,A_l\right)^{-1}~,~\quad
  \quad\quad\quad\quad\quad\quad\quad\quad\quad\quad\quad\quad\quad\quad\quad\quad\quad\quad\quad\quad\quad\quad
   ~\\
   \nonumber\\
 {\cal{T}}_{\textstyle{_{lmpq}}}\sim R^{\,2l+1}k_l\,\sin\epsilon_l\sim\frac{R^{\,2l+1}\,A_l}{\left(1+A_l\right)^2}\sim
 \left\{
 \begin{array}{c}
 R^{\,2l-1}~,~~\mbox{for}~~ A_l\ll 1~~\mbox{(superearths)}~,~\quad\quad\quad\quad\quad\quad\quad\quad\quad\quad\quad\\
 ~\\
 R^{\,2l+3}~,~~\mbox{for}~~ A_l\gg 1~~\mbox{(small
 bodies, ~small terrestrial planets)}~.
 \end{array}
 \right.
 \label{cont}
 \ea\\

 In the closest vicinity of the $\,lmpq\,$ commensurability, i.e., when the tidal frequency $\,\chi_{\textstyle{_{lmpq}}}\,$ approaches
 zero, the tidal factors's behaviour is described by (\ref{cccc}). This furnishes a different scaling law for the torque, and the form
 of this law is the same for telluric bodies of all sizes:
 \ba
 \nonumber
 \mbox{Over the frequency band}\quad\chi\,\ll\,\tau^{-1}_{_M}\,\left(1\,+\,A_l\right)^{-1}~,~\quad\quad\quad\quad\quad\quad\quad\quad\quad\quad~
 \quad\quad\quad\quad\quad\quad\quad\quad~
 \quad\quad
 ~\\
 \nonumber\\
 \quad\quad
 {\cal{T}}_{\textstyle{_{lmpq}}}\sim R^{\,2l+1}k_l\,\sin\epsilon_l\sim R^{\,2l+1}\,A_l\sim\,R^{\,2l-1}~~~.\quad\quad\quad
 \quad\quad\quad\quad\quad\quad\quad\quad\quad\quad\quad\quad\quad\quad\quad
 \label{contr}
 \ea

 \section{Conclusions and examples}\label{subsection}


 Within the inelasticity-dominated band, the phase lags in a homogeneous near-spherical body and in a sample of material interrelate as
 \ba
 \tan\epsilon_{\textstyle{_l}}\,=\,\frac{A_l}{1\,+\,A_l}~\tan\delta\,~\approx~
 \left\{
 \begin{array}{c}
 A_l~\tan\delta\quad\quad\mbox{for}\quad A_l\ll 1~~
 \mbox{(superearths)}~~,\quad\quad\quad\quad\quad\quad\quad\quad\quad\quad\quad\\
 ~\\
 \quad\quad\tan\delta\quad~~~\mbox{for}\quad A_l\gg 1~~\mbox{(small
 bodies, ~small terrestrial planets)}~\,.
 \end{array}
 \right.
 \label{Saint}
 \ea
 However within the transitional zone, the link between the seismic and tidal dissipation rates becomes more complicated.

 The interrelation between the tidal and seismic damping becomes apparently paradoxical at low frequencies, where viscosity dominates.
 As can be seen from (\ref{90} - \ref{sup}), in the zero-frequency limit the tidal and seismic $\,Q\,s\,$ (if defined as cotangents of
 the appropriate lags) become {\it{inversely}} proportional to one another:
 \ba
 \tan\epsilon_{\textstyle{_l}}\,~\approx~A_l\,\chi\,\tau_{_M}~=~\frac{\textstyle A_l}{\,\textstyle \tan\delta\,}~\quad\quad
 \quad\quad\mbox{for}\quad\quad 0\leq\chi\ll\frac{\textstyle 1}{\textstyle{\tau_{_M}}\,\sqrt{A_{\textstyle{_l}}+1}}~~~.\,~~~
 \label{paradox}
 \ea
 This behaviour however has a good qualitative explanation -- the average tidal torque $\,lmpq\,$ should vanish on crossing of the $\,lmpq\,$
 resonance.

 While in qualitative discussions it is easier to deal with the quality factors $\,Q_l\,$, in practical calculations we should rely on
 the factors $\,k_l\,\sin\epsilon_l\,$, which show up in the Darwin-Kaula expansion of tides. Just as $\,\tan\epsilon_l\,$, so the
 quantity $\,k_l\,\sin\epsilon_l\,$ too becomes linear in $\,\chi\,$ for low values of $\,\chi\,$. As we saw in subsection \ref{be},
 this happens over the frequencies below $\,\chi\ll\frac{\textstyle 1}{\textstyle \tau_{_M}\,{(A_l+1)\,}}~$. The slight difference
 between this threshold and the one shown in (\ref{paradox}) stems from the fact that not only the lag but also the Love number is
 frequency dependent.

 The factors $\,k_l\,\sin\epsilon_l\,$ bear dependence upon the radius $\,R\,$ of a tidally disturbed primary, and the form of this
 dependence is not always trivial. At low frequencies, this dependence
 follows the intuitively obvious rule that the heavier the body the stronger it mitigates tides (and thence the smaller the value of
 $\,k_l\,\sin\epsilon_l\,$). However at high frequencies the calculated frequency dependence obeys this rule only beginning from
 sizes about or larger than the double size of the Earth, i.e., when self-gravitation clearly plays a larger role in tidal friction
 than the rheology does -- see the discussion at the end of subsection \ref{last}.

 The dependence of $\,k_l\,\sin\epsilon_l\,$ upon $\,R\,$ helps one to write down the overall $\,R$-dependence of the tidal torque.
 Contrary to the common belief, the $\,lmpq\,$ component of the torque does {\it{not}} scale as $\,R^{\,2l+1}\,$, see formulae
 (\ref{cont}) and (\ref{contr}).\\

 Here follow some examples illustrating how our machinery applies to various celestial bodies.
 \begin{itemize}

 \item[\bf 1.~] For small bodies and small terrestrial planets, the effect of self-gravitation is negligible, except in the closest
 vicinity of the zero frequency. Accordingly, for these bodies there is no difference between the tidal and seismic
 dissipations.$\,$\footnote{~This can be understood also through the following line of reasoning. For small objects, we have $\,A_{
 \textstyle{_l}}\gg 1\,$; so the complex Love numbers (\ref{k2bar}) may be approximated with
 \ba
 \nonumber
 \bar{k}_{\textstyle{_l}}(\chi)\,=\;-\;\frac{3}{2}\;\frac{{\textstyle\,\stackrel{\mbox{\bf \it \_}}{J}
 (\chi)}}{{\textstyle\,\stackrel{\mbox{\bf \it \_}}{J}
 (\chi)}\;+\;A_l\;{\textstyle J}}\;=\;-\;\frac{3}{2}\;\frac{{\textstyle\,\stackrel{\mbox{\bf \it \_}}{J}
 (\chi)}}{A_l\;{\textstyle J}}~+~O\left(~|{\textstyle\,\stackrel{\mbox{\bf \it \_}}{J}
 }/(A_l\,J)\,|^2~\right)\;\;\;.~~~~~
 \label{pivotal}
 \ea
 The latter entails:
 \ba
 \nonumber
 \tan\epsilon_{\textstyle{_l}}(\chi)\;\equiv\;
 -\;\frac{{\cal{I}}{\it{m}}\left[\bar{k}_{\it{l}}(\chi)\right]}{{\cal{R}}{\it{e}}
 \left[\bar{k}_{\it{l}}(\chi)\right]}\;\approx\;
 -\;\frac{{\cal{I}}{\it{m}}\left[\bar{J}(\chi)\right]}{{\cal{R}}{\it{e}}\left[\bar{J}(\chi)\right]}
 \;=\;\tan\delta(\chi)\;\;\;,
 \label{epsidelta}
 \ea
 which is, in fact, correct {\it{up to a sign}} -- see the closing paragraph of subsection \ref{4.1}.}

 Things change in the closest vicinity of the zero frequency. As can be observed from the second line of (\ref{90}), for small bodies
 and small planets the tangent of the tidal lag
 becomes linear in the tidal frequency $\,\chi\,$ when the frequency $\,\chi\,$ becomes short of a certain threshold:\footnote{~Recall
 that for small objects $\,A_l\gg 1\,$.}
 $~\,\chi\ll\frac{\textstyle 1}{\textstyle \tau_{_M}\,\sqrt{A_l+1\,}}\,\approx\,\frac{\textstyle 1}{\textstyle\tau_{_M}\,\sqrt{A_l}}~$.
 As can be seen from (\ref{cccc}), the tidal factor $\,k_l\,\sin\epsilon_l\equiv|\bar{k}_{l}(\chi)|\,\sin\epsilon_l(\chi)\,$ becomes linear in $\,\chi\,$ for
 $~\,\chi\ll\textstyle \tau_{_M}^{-1}\,(A_l+1)^{-1}\,\approx\,\textstyle\tau_{_M}^{-1}\,A_l^{-1}~$.
 ~\\

 \item[\bf 2.~] Tidal dissipation in superearths is much less efficient than in smaller terrestrial planets or moons -- a
 circumstance that should reduce considerably the rates of orbit circularisation. This cautionary point has ramifications also upon
 the other tidal-dynamic timescales (e.g., despinning, migration).

 In simple words, self-gravity reduces tidal dissipation because gravitational attraction pulls the tidal bulge back down, and thus
 reduces strain in a way similar to material strength.

 As can be seen from (\ref{sup}), at tidal frequencies $\,\chi\,$ lower than the inverse Maxwell time,\footnote{~For superearths,
 $\,A_l\ll 1\,$.} the tangent of the tidal lag changes its behaviour considerably, thereby avoiding divergence at the zero frequency.
 According to (\ref{cccc}), the same pertains to the factor $\,k_l\,\sin\epsilon_l\,$.
 ~\\

 \item[\bf 3.~] While the role of self-gravity is negligible for small planets and is dominant for superearths, the case of the Earth
 is intermediate. For our mother planet, the contribution of self-gravitation into the Love numbers and phase lags is noticeable,
 though probably not leading. Indeed, for $\,\mu\approx 0.8\times 10^{11}~$Pa, one arrives at:
 \ba
 A_{2}\,\approx~2.2~~~,
 \label{}
 \ea
 so formula (\ref{Earth}) tells us that the Earth's tidal quality factor is a bit larger than its seismic counterpart, {\it{taken at
 the same frequency}}:$\,$\footnote{~When Benjamin et al. (2006) say that, according to their data, the tidal quality factor is slightly
 lower than the seismic one, these authors compare the two $\,Q\,$ factors measured at different frequencies. Hence their statement is in
 no contradiction to our conclusions.}
 \ba
 \label{Earth_2}
 {}^{(tidal)}Q_{\textstyle{_2}}^{(solid~Earth)}\,\approx~1.5~\times~\,{}^{(seismic)}Q^{(solid~Earth)}
 {\left.~~\right.}_{\textstyle{_{\textstyle{.}}}}
 \ea
 The geodetic measurements of semidiurnal tides, carried out by Ray et al. (2001), yield $\,{}^{(tidal)}Q_{\textstyle{_2}}^{(solid~Earth)}
 \,\approx\,280\,$. The seismic quality factor $\,{}^{(seismic)}Q^{(solid~Earth)}\,$ varies over the mantle, assuming values from
 $100$ through $300$. Accepting $200$ for an arguable average, we see that (\ref{Earth_2}) furnishes a satisfactory qualitative
 estimate.

 This close hit should not of course be accepted too literally, taken the Earth's complex structure and the uncertainty in our knowledge
 of the Earth's rigidity. Still, on a qualitative level, we may enjoy this proximity with cautious optimism.\\

 \item[\bf 4.~] The case of the Moon deserves a special attention. Fitting of the LLR data to the power scaling law $\,Q\sim\chi^{p}\,$ has
 rendered a small {\it{negative}} value of the exponential: $\,p\,=\,-\,0.19~$ (Williams et al. 2001). Further attempts by the JPL team to
 reprocess the data have led to $\,p\,=\,-\,0.07~$. $\,$According to Williams \& Boggs (2009),\\
 ~\\
 ``{\it{There is a weak dependence of tidal specific dissipation $\,Q\,$ on period. The $\,Q\,$ increases from $\,\sim 30\,$ at a
 month to $\,\sim 35\,$ at one year. $~Q\,$ for rock is expected to have a weak dependence on tidal period, but it is expected to
 decrease with period rather than increase. The frequency dependence of $\,Q\,$ deserves further attention and should be improved.}}"
 \vspace{3mm}

 To understand the origin of the small negative value of the power, recall that it emerged through fitting of the tidal $Q_2$ and not
 of the seismic $Q$. If the future laser ranging confirms these data, this will mean that the principal tide in the Moon is located
 close to the maximum of the inverse tidal quality factor, i.e., close to the maximum taken by $\,\tan\epsilon_2\,$ in (\ref{90}) at
 the frequency inverse to $\,\tau_{_M}\,\sqrt{A_l}\,$. Rigorously speaking, it was of course the factor $\,k_2\,\sin\epsilon_2\,$ which
 was actually observed. The maximum of this factor is attained at the frequency $\,\tau_{_M}^{-1}\,
 (A_{\textstyle{_l}}+1)^{-1}\,$, as can be seen from (\ref{bbbb} - \ref{cccc}). It then follows from the LLR data
 that the corresponding timescale $\,\tau_{_M}\,(A_{\textstyle{_l}}+1)\,$ should be of order $\,0.1\,$ year. As explained in Efroimsky (2012), this would set the mean viscosity of the Moon
 as low as
 \ba
 \eta_{\textstyle_{Moon}}\,=~3\,\times\,10^{16}~\mbox{Pa~s}~~~,
 \label{}
 \ea
 which in its turn would imply a very high concentration of the partial melt in the low mantle -- quite in accordance with the existing
 models (Nakamura et al. 1974, Weber et al. 2011).

 The future LLR programs may be instrumental in resolving this difficult issue. The value of the exponential $\,p\,$ will have
 ramifications for the current models of the lunar mantle.

 \end{itemize}

 \section{Comparison of our result with that by Goldreich (1963)}

 A formula coinciding with our (\ref{Earth}) was obtained, through remarkably economic and elegant semi-qualitative reasoning, by Peter
 Goldreich (1963).

 The starting point in {\it{Ibid.}} was the observation that the peak work performed by the second-harmonic disturbing potential should
 be proportional to this potential taken at the primary's surface, multiplied by the maximal surface inequality:
 \ba
 E_{peak}\,\sim~R^5\,~\frac{R}{\frac{\,\textstyle 19\,\mu\,}{\,\textstyle  2\,\mbox{g}\,\rho\,R}\,+\,1\,}~\sim~\frac{R^7}{19\,\mu\,+\,2\,\mbox{g}\,\rho\,R}~~~,
 \label{one}
 \ea
 $R\,$ being the primary's radius.

 In the static theory of Love, the surface strain is proportional to $\,R^2/\left(\,19\,\mu\,+\,2\,\mbox{g}\,\rho\,R\,\right)\,$. The
 energy loss over a cycle must be proportional to the square of the surface strain. Integration over the volume will give an extra
 multiplier of $\,R^3\,$, up to a numerical factor:
 \ba
 \Delta E_{cycle}\,\sim~-~\frac{R^7}{\left(\,19\,\mu\,+\,2\,\mbox{g}\,\rho\,R\,\right)^2}~~~.
 \label{two}
 \ea
 Comparison of (\ref{one}) and (\ref{two}) rendered
 \ba
 Q~=~-~\frac{2\pi~E_{peak}}{\Delta E_{cycle}}~\sim~\left(\,19\,\mu\,+\,2\,\mbox{g}\,\rho\,R\,\right)~~~,
 \nonumber
 \ea
 wherefrom Goldreich (1963) deduced that
 \ba
 \frac{Q}{Q_0}~=~1~+~\frac{2\,\mbox{g}\,\rho\,R}{19\,\mu}~~~,
 \nonumber
 \ea
 $Q_0\,$ being the value of $\,Q\,$ for a body where self-gravitation is negligible. This coincides with our formula (\ref{Earth}).

 In reality, the coincidence of our results is only partial, for two reasons:

 \begin{itemize}

 \item{} First, our derivation of the right-hand side of (\ref{kirha_1}) was based on the prior convention that the quantity $\,J\,$
 entering expression (\ref{new}) is the {\it{unrelaxed}} compliance $\,J(0)\,$ of the mantle. Accordingly, the quantity $\,\mu=1/
 J\,$ entering the expression for $\,A_l\,$ should be the {\it{unrelaxed}} rigidity $\,\mu(0)=1/J(0)\,$. In Goldreich (1963) however,
 the static, i.e., {\it{relaxed}} moduli were implied.

 In {\it{Ibid.}}, this mismatch was tolerable, because the paper was devoted to small bodies. For these objects, $\,A_l\,$ is large,
 no matter whether we plug the relaxed or unrelaxed $\,\mu\,$ into (\ref{L4}). Thence the difference between the tidal and seismic
 $\,Q\,$ factors is small, as can be seen from the second line of (\ref{Saint}).

 For earths and superearths, however, the distinction between the unrelaxed and relaxed (static) moduli is critical. As can be seen
 from the first line of (\ref{Saint}), the tidal $\,Q\,$ factor is inversely proportional to $\,A_l\,$ and, thereby, is
 inversely proportional to the mantle rigidity $\,\mu\,$. As well known (e.g., Ricard et al. 2009, Figure 3), the unrelaxed $\,\mu
 \,$ of the mantle exceeds the relaxed $\,\mu\,$ by about two orders of magnitude.

 \item{} Second, as our calculation demonstrates, the simple interrelation given by (\ref{Earth}) and (\ref{Saint}) works {\it{only
 in the inelasticity-dominated band}}. In the transition zone (which begins, in the solid Earth, at timescales longer than $\sim$ 1 yr)
 and in the viscosity-dominated band of lower frequencies, the interrelation between the tidal and seismic lagging is more complicated,
 and it deviates from Goldreich's formula in a fundamental way. In the zero-frequency limit, the cleavage between the tidal and
 seismic dissipation laws gets even larger: the tidal and seismic Q$\,$s become not proportional but {\it{inversely}} proportional
 to one another. Description of tidal lagging in all these, low-frequency bands requires a rheological model and the subsequent
 mathematics, and cannot be obtained through the simple arguments used by Goldreich (1963).


 \end{itemize}

 \noindent
 Despite these differences, the estimate by Goldreich (1963) provided as close a hit as was possible without resorting to heavy
 mathematics. The elegance of Peter Goldreich's arguments and the depth of his insight are especially impressive, taken the complexity
 of the problem and the volume of calculations required to obtain the exact answer.\\

 ~\\
 {\Large{\bf Acknowledgements}}\\

 To a large extent, my understanding of the theory of bodily tides was developing through the enlightening conversations which I had
 on numerous occasions with Bruce Bills, Julie Castillo-Rogez, V\'eronique Dehant, Sylvio Ferraz-Mello, Val\'ery Lainey, Valeri
 Makarov, Francis Nimmo, Stan Peale, Tim Van Hoolst, and James G. Williams. It is a great pleasure for me to thank deeply all these
 colleagues for their time and advise. Needless to say, none of them shares the responsibility for my possible omissions.

 I also wish to pay tribute to the late Vladimir Churkin, whose tragic death prevented him from publishing his preprint cited in this
 paper. Written with a great pedagogical mastership, the preprint helped me to understand how the Love-number formalism should be
 combined with rheology.

 My special gratitude is due to Shun-ichiro Karato for the help he so kindly provided to me, when I was just opening for myself this
 intriguing area, and for the stimulating exchanges, which we have had for years since then.

 Last, and by no means least, I sincerely appreciate the support from my colleagues at the US Naval Observatory, especially from John
 Bangert.
 \noindent
 \section*{}
\noindent
\begin{table*}[htdp]
{\underline{\textbf{\Large{Appendix}}}}
~\\
~\\{{\textbf{\Large{Symbol Key}}}}
\begin{center}
{\scriptsize
\begin{tabular}{lll}
\noalign{\smallskip}
\hline
\hline\\
$ A_l $ & Dimensionless product emerging in the denominator of the expression for the Love number $\,k_l\,$  \\
$ E $ & Energy \\
$ {\cal{E}} $ & Empirical constant having the dimensions of time, in the generic rheological law (\ref{generic}) \\
  g   & Surface gravity \\
$ G $ & Newton's gravitational constant \\
$ l $ & Degree (spherical harmonics, Legendre polynomials) \\
$ m $ & Order (spherical harmonics, associated Legendre polynomials) \\
$ J, ~J(0) $ & Unrelaxed compliance \\
$ J(\infty) $ & Relaxed compliance \\
$ J(t\,-\,t\,') $ & Creep-response function (compliance function, kernel of the compliance operator) \\
$ \hat{J} $ & Compliance operator \\
$ k_l $ & Tidal Love number of degree $l$  \\
$ k_l(t-t\,') $ & Kernel of the Love operator of degree $l$  \\
$ \hat{k}_l $ & The Love operator of degree $l$  \\
$ \bar{k}_l(\chi) $ & Fourier component, at frequency $\chi$, of the time derivative of the kernel $k_l(t-t\,')$  \\
     ${\cal{M}}$ & Mean anomaly \\
     $ n $ & Mean motion \\
$ p $ & Exponential in the generic rheological law (\ref{generic}) \\
$ P_l $ & Legendre polynomials of degree $l$ \\
$ P_{lm} $ & Legendre associated functions (associated Legendre polynomials) of degree $l$ and order $m$ \\
$ Q $ & Dissipation quality Factor \\
$ r$ & Distance \\
 $ {\erbold} $ & Vector connecting the centre of the tidally-perturbed body (interpreted as the primary) with a point exterior to this body\\
 $ {\erbold}^{\,*} $ & Vector connecting the centre of the tidally-perturbed body (the primary) with a point-like tide-raising secondary\\
$ R $ & Primary's mean radius\\
$ t $ & Time \\
$ u_{\gamma\nu} $ & Shear strain tensor\\
$ \bar{u}_{\gamma\nu}(\chi) $ & Fourier component, at frequency $\,\chi\,$, of the shear strain tensor\\
$ U $ & Change in the potential of the tidally-perturbed body (interpreted as the primary)\\
$ W $ & Disturbing potential generated by the tide-raising body (interpreted as the secondary)\\
$ \alpha, \beta $ & Parameters of the Andrade model \\
$ \gamma, \nu $ & Tensor indices \\
$ \Gamma $ & the Gamma function \\
$ \delta $ & Material phase lag \\
$ \Delta t $ & Time lag \\
$ \epsilon $ & Tidal phase lag  \\
$ \epsilon_{\textstyle{_{lmpq}}} $ & Tidal phase lag of the mode $lmpq$ in the Darwin-Kaula expansion\\
 $ \lambda $ & Longitude \\
$ \zeta $ & Parameter of the reformulated Andrade model ~(ratio of the Andrade timescale $\tau_{_A}$ to the Maxwell time $\tau_{_M}$)\\
$ \eta $ & Viscosity \\
$ \mu, ~\mu(0) $ & Unrelaxed shear modulus (unrelaxed rigidity) \\
$ \mu(\infty) $ & Relaxed shear modulus (relaxed rigidity) \\
$ \mu(t-t\,') $ & Stress-relaxation function (kernel of the rigidity operator) \\
$ \hat{\mu} $ & Rigidity operator\\
 $ \phi $ & Latitude \\
$ \rho $ & Mass density \\
$ \sigma_{\gamma\nu} $ & Shear stress tensor\\
$ \bar{\sigma}_{\gamma\nu}(\chi) $ & Fourier component, at frequency $\,\chi\,$, of the shear stress tensor\\
$ \tau $ & Time \\
$ \tau_{_M} $ & Maxwell time (viscoelastic timescale) \\
$ \tau_{_A} $ & Andrade time (inelastic timescale) \\
$ {\cal{T}} $ & Tidal torque \\
$ \Theta(t-t\,') $ & the Heaviside function \\
$ \theta $ & Sidereal angle of the tidally-disturbed body (interpreted as the the primary)\\
$ \stackrel{\bf\centerdot}{\theta\,} $ & Spin rate of the primary\\
$ \varphi_\sigma\,,~\varphi_u $ & Initial phases of the stress and strain \\
$ \chi $ & Frequency\\
$ \chi_{_0} $ & Frequency threshold marking the boundary between the inelasticity- and viscosity-dominated frequency bands\\
$ \chi_{\textstyle{_{lmpq}}} $ & Physical frequencies of deformation emerging in the tidal theory (absolute values of the tidal modes
   $\omega_{\textstyle{_{lmpq}}}$ )\\
$ \omega_{\textstyle{_{lmpq}}} $ & Tidal modes in the Darwin-Kaula expansion of tides\\
$ \omega $ & Argument of the pericentre \\
$ \Omega $ & Longitude of the node\\
\hline
\hline
\end{tabular}
}
\end{center}
\label{tab:symb}
\end{table*}

 \pagebreak

 \section*{}
 {}

 \end{document}